%% file: paper3d.tex
\documentclass[iop, numberedappendix]{emulateapj}
\shorttitle{Non-ideal Gravitational instability}
\shortauthors{M.-K.\ Lin \& K. M. Kratter}

\usepackage{amsmath}
\usepackage{paralist}
\usepackage{bm}
\newcommand{\p}{\partial}
\newcommand{\zmax}{z_\mathrm{max}}

\newcommand{\ii}{\mathrm{i}}

\newcommand{\qthree}{Q_\mathrm{3D}}
\newcommand{\qtwo}{Q}

\newcommand{\tcool}{t_\mathrm{c}}
\newcommand{\tirr}{T_\mathrm{irr}}

\newcommand{\yr}{\mathrm{yr}}
\newcommand{\qvert}{Q_\mathrm{3D}}
\defcitealias{lin12}{L12} 
\defcitealias{nelson13}{N13}
\defcitealias{barker15}{BL15}
\defcitealias{mcnally14}{MP14}
\defcitealias{goldreich67}{GS67}
\defcitealias{rafikov15}{R15}

\begin{document}

\title{On the Gravitational stability of gravito-turbulent accretion disks}
\author{Min-Kai Lin\altaffilmark{1} \& Kaitlin M. Kratter}
\affil{Department of Astronomy and Steward Observatory, \\
University of
  Arizona, 933 North Cherry Avenue, Tucson, AZ 85721, USA}
\altaffiltext{1}{Steward Theory Fellow}
\email{minkailin@email.arizona.edu}

\begin{abstract}
  Low mass, self-gravitating accretion disks admit quasi-steady,
  `gravito-turbulent' states in which cooling balances turbulent viscous 
  heating. However, numerical simulations show that gravito-turbulence
  cannot be sustained beyond dynamical timescales when the cooling
  rate or corresponding turbulent viscosity is too large. The result is disk fragmentation.         
  We motivate and quantify an interpretation of disk 
  fragmentation as the inability to maintain gravito-turbulence due to  
  formal secondary instabilities driven by:  
\begin{inparaenum}[1)] 
\item 
  cooling, which reduces pressure support; and/or
\item 
  viscosity, which reduces rotational support. 
\end{inparaenum}
We analyze the axisymmetric gravitational stability of viscous, non-adiabatic
accretion disks with internal heating, external irradiation, and
cooling  in the shearing box approximation.   
We consider parameterized cooling functions in 2D and 3D
disks, 
as well as   
radiative diffusion in 3D. We show that 
generally there is no critical cooling rate/viscosity below which 
the disk is formally stable, although interesting limits appear 
for unstable modes with lengthscales on the order of the disk thickness.   
We apply this new linear theory to protoplanetary disks subject to
gravito-turbulence modeled as an effective 
viscosity, and cooling regulated by dust opacity. 
We find that viscosity renders the disk beyond $\sim 60$AU dynamically
unstable on radial 
lengthscales a few times the local disk thickness. This is coincident 
with the empirical condition for disk fragmentation based on a
maximum sustainable stress. 
We suggest turbulent stresses can play an active role in realistic disk 
fragmentation by removing rotational stabilization against 
self-gravity, and that the observed transition in behavior from 
gravito-turbulent to fragmenting may reflect    
instability of the gravito-turbulent state itself.
\end{abstract}

\section{Introduction}\label{intro} 
Understanding the gravitational stability of rotating disks    
is central to many astrophysical problems \citep{kratter16}. In the context of 
gaseous protostellar or protoplanetary disks (PPDs), gravitational instability (GI)
has two applications. It can provide gravitational torques to
transport angular momentum outwards and thus enable mass accretion
\citep{armitage10,turner14}. GI may also lead to disk fragmentation,
which has been invoked to explain the formation of stellar/sub-stellar companions or 
giant planets 
at large radii \citep{boss97,kratter06,stam09,helled14} 
Studying these non-linear phenomena requires
direct numerical simulations. Nevertheless, physical insight can be    
obtained through analytical modeling. 

The standard metric for the (inverse) strength of disk self-gravity is
the \emph{Toomre parameter}, 
\begin{align}\label{toomreQ_criterion}
  Q \equiv \frac{c_s\kappa}{\pi G \Sigma}  
\end{align}
\citep{toomre64}. Here, $c_s$ is the isothermal sound-speed,
$\kappa$ is the epicyclic frequency (which equals the rotation
frequency $\Omega$ in a Keplerian disk), $\Sigma$ is the surface
density and $G$ is the gravitational constant. The Toomre parameter is a
measure of the destabilizing effect of self-gravity ($G\Sigma$) against 
the stabilizing effect of rotation ($\kappa$) and pressure 
($c_s$). This is evident from the dispersion relation, 
\begin{align}\label{classic_dr}
  s^2 =  2\pi G\Sigma|k| - \kappa^2 - c_s^2k^2,   
\end{align}
which relates the growth rate $s$ and radial wavenumber $k$ for local,
axisymmetric waves in a two-dimensional (2D, razor-thin), inviscid and
isothermal\footnote{Eq. \ref{toomreQ_criterion}---\ref{classic_dr} also apply to adiabatic disks if one takes $c_s$ as the adiabatic sound-speed.} disk.  
When $Q<1$, there is a range of $k$ for which such disturbances are
unstable.  
Non-axisymmetric modes can develop for larger, but still order-unity
values of $Q$ 
\citep{lau78,papaloizou89,papaloizou91}. 

We emphasize that the oft-used dispersion relation and corresponding
Toomre parameter (Eqs. \ref{toomreQ_criterion} and \ref{classic_dr})
are derived from \emph{idealized} conditions:  
the base disk is laminar, inviscid, and does not experience any net thermal 
losses\footnote{Isothermal disks implicitly assume  
  heating and thermal losses are exactly balanced at all 
  times, so these do not cool in the current context.} 
(hereafter `cooling').  
However, real disks can cool, for example, 
due to radiative losses. Accretion disks may also be turbulent, the 
dynamic and thermodynamic effect of which is often modeled through an 
effective viscosity \citep{shakura73,lin87,armitage01,rafikov15}. 

However, the Toomre parameter  
is still widely applied to viscous (turbulent), cooling
accretion disk models \citep[e.g.][]{gammie01,cossins09, kimura12}, 
despite the mismatch in the included underlying physics. 
We show that including non-ideal physics, such as cooling and
viscosity, in fact modifies the classic dispersion 
relation, and hence the condition for GI. Thus the Toomre condition
alone is insufficient to assess GI in realistic disks.  

The goal of this work is to generalize the analytic treatment of disk
GI, by including cooling and viscosity, to allow 
a self-consistent discussion of GI in realistic accretion disk
models.  
We first review in \S\ref{cool_gi} and \S\ref{visc_gi}, respectively,  
how cooling or viscosity can lead to GI even when $Q>1$, by removing
pressure or rotational support. We then 
discuss in \S\ref{frag_intro}---\S\ref{gi_of_gi} how these
effects may relate to the transition from self-regulated GI to
fragmentation seen in numerical simulations. The rest of this paper is
laid out in \S\ref{plan}.

\subsection{Cooling-driven gravitational instability}\label{cool_gi}
Cooling reduces pressure support against self-gravity. The
cooling time  
$\tcool$ is the timescale over which the disk temperature $T$ is relaxed
to some floor value, which may be zero. 
 It is often written as 
\begin{align} \label{beta_def}
  \tcool = \beta \Omega^{-1},
\end{align}
where $\beta$ is the corresponding dimensionless cooling time. This
type of parameterized cooling, first applied by \cite{gammie01},
allows a range of thermodynamic responses to be explored. 

In reality, PPD cooling is controlled by radiation   
from dust grains \citep{bell94,dalessio97,chiang97}. Although most
PPDs subject to GI are optically thick, cooling parameterized by the
$\beta$ model formally only captures optically thin cooling when
used in numerical simulations. It is, however, possible to modify the
standard cooling function to mimic optically-thick cooling (see,
e.g. \S\ref{2dppd}).  

Previous work has quantified the role of cooling primarily as a means to reduce the sound speed term
in $Q$ \citep[though see e.g.,][]{clarke07}. Here we will quantify how cooling enables GI
even when $Q>1$. For example, if perturbations can
cool to arbitrarily low temperatures (which, in fact, is a common
cooling prescription in numerical simulations), we find the above
dispersion relation is  
modified to read 
\begin{align}\label{cool_dr}
s^2 = 2\pi G\Sigma |k| - \kappa^2 - \gamma \left(\frac{\tcool s}{1 +
  \tcool s}\right)c_s^2k^2,
\end{align}
where $\gamma$ is the adiabatic index. Cooling increases the growth
rate by reducing the magnitude of the 
pressure term. In fact, for $t_c\in[0,\infty)$ a formal condition
for instability is  
\begin{align*}
  2\pi G \Sigma |k| > \kappa^2,
\end{align*}
which is what would be obtained from Eq. \ref{classic_dr} with 
pressure neglected ($c_s\to 0$). This condition does not actually
depend on the cooling time, and instability is
possible for any finite $Q$. 
Thus, the mere presence of cooling changes the qualitative nature of
GI compared to the simple Toomre condition. 
   


\subsection{Viscosity-driven gravitational instability}\label{visc_gi}
Viscous disks can also develop GI even when $Q>1$. This is because, as 
demonstrated below, viscosity removes rotational support against  
self-gravity for long-wavelength disturbances 
\citep{lynden-bell74,willerding92,gammie96}. A similar effect   
occurs in dusty fluids where the required frictional forces are 
provided by dust-gas drag \citep{goodman00,ward00, takahashi14}. 
In fact, this
is a mechanism to enhance particle clumping for planetesimal
formation 
 \citep{youdin05,youdin11}. Therefore it is not unreasonable to expect analogous
fragmentation in gaseous disks due to viscosity.   

It is conventional to write the kinematic viscosity $\nu$ as   
\begin{align}\label{alpha_param}
  \nu  = \alpha \frac{c_s^2}{\Omega}, 
\end{align}
where $\alpha$ is the dimensionless viscosity
coefficient \cite{shakura73}. This
parameterization can be modified to include more complex dependencies on the
fluid variables, but Eq. \ref{alpha_param} is the general form. 

For an isothermal, viscous, self-gravitating disk in 2D,
\cite{gammie96} finds the approximate dispersion relation   
\begin{align}
s \simeq \frac{\nu k^2\left(2\pi G\Sigma |k| -
  c_s^2k^2\right)}{\kappa^2 + c_s^2k^2 - 2\pi G\Sigma|k|}. 
\end{align}
Assuming $Q>1$, instability occurs if 
\begin{align*}
  2\pi G \Sigma > c_s^2|k|,
\end{align*}
provided that $\nu k^2\neq0$. This condition for viscous GI is 
identical to what would be obtained from Eq. \ref{classic_dr} with
rotation neglected ($\Omega\to 0$). That is, a classically stable
disk can be destabilized by viscosity as it reduces 
rotational stabilization \citep{lynden-bell74}.  

For gaseous accretion disks, a Navier-Stokes viscosity is often
implemented (as above) to mimic hydrodynamic or magneto-hydrodynamic
(MHD) turbulence \citep{shakura73}. How well turbulence can be modeled
as an effective viscosity is a separate issue 
\citep{balbus99}. However, it is reasonable to assume that these
mechanisms, which are observed in simulations  
to transport angular momentum outwards (and mass inwards),
frustrate rotational support. 

In this work, we use viscosity to model two possible
physical effects of turbulence: heating via dissipation and angular momentum transport. 
We emphasize that our model for GI enabled by viscosity (hereafter
viscous GI) does not assume a particular origin for the viscosity. GI
itself or other forms of magnetic or hydrodynamic turbulence are 
allowed in this framework. We will generalize the theory of viscous GI
to include an energy equation  with viscous heating, irradiation,
explicit cooling, as well as three-dimensionality (3D) in order to
consider viscous GI in more realistic PPD models.   


\subsection{Relevance to gravito-turbulent disk fragmentation}\label{frag_intro}
We develop a general framework for viscous disks without  
assuming a specific origin for the viscosity. 
We will, however, apply the theory to `gravito-turbulent'
disks in which the viscosity is associated with some underlying
(classic) GI, as described below.  

Consider an initially laminar, $Q\gg1$ disk, without external 
heating, as it cools. The disk temperature will decline
until $Q=O(1)$, whence non-axisymmetric modes grow and heat the disk
through the dissipation of spiral shocks \citep{cossins09}. 
This setup permits a quasi-steady, turbulent state with
$Q =O(1)$ in which cooling is balanced by shock heating to maintain
thermal equilibrium \citep{gammie01,shi14}. 

  Global disk simulations \citep[e.g.][]{lodato04} show
  that the transport and 
  heating associated with this gravito-turbulence may be described as a
  local viscous process provided that the disk-to-star mass ratio is
  small ($\lesssim0.25$) and the disk is thin (aspect-ratio $\lesssim
  0.1$).  

 In this case the $\alpha$ and $\beta$ parameters defined above
are inversely related (see, 
e.g. Eq. \ref{alpha_beta_relation}). Recent 
  vertically-extended shearing box simulations 
  also confirm this relation in 3D disks \citep{shi14}. 
   
Numerical experiments, however, show that if the cooling time is too
small (or the viscosity is too large), say,   
\begin{align}\label{frag_cond}
  \beta < \beta_\mathrm{c} \quad
  \left(\alpha>\alpha_\mathrm{c}\right), 
\end{align}
then the disk fragments \citep{gammie01,rice05,rice11}. In fact, in 
the absence of global effects, numerical simulations show that there are
two --- and only two --- possible outcomes for self-gravitating,
cooling disks: gravito-turbulence or fragmentation. 

There is considerable debate on the exact value of $\beta_\mathrm{c}$  
and whether or not a critical cooling time can be defined at all
\citep{meru11,lodato11,meru12,paardekooper12,hopkins13}. There are, in
addition, numerical convergence issues when simulating disk  
fragmentation, which we discuss in \S\ref{summary}. 

However, it is generally accepted that \emph{steady, gravito-turbulent
  disks do not  
  exist for sufficiently rapid cooling or large viscosity} \citep{johnson03}.   
This is intriguing because, as highlighted in
\S\ref{cool_gi}---\ref{visc_gi}, cooling or viscosity can reduce 
gravitational stability independently of their influence on the exact
value of the classic $Q$. This 
motivates a physical      
interpretation of disk fragmentation as the inability to maintain a
gravito-turbulent state due to 
secondary instabilities driven by cooling and/or viscosity.

\subsection{Instability of the gravito-turbulent state}\label{gi_of_gi}
  In this work, we formally treat the gravito-turbulent
  disk described above as an equilibrium state, to which we apply    
  standard linear stability analysis. Thus by `perturbations' we mean   
  deviations away from this gravito-turbulent 
  basic state\footnote{This should not be confused with 
    fluctuations with respect to the 
    laminar disk that maintain the underlying
    gravito-turbulence.}. 
  In our model, we interpret instability of such perturbations  to 
  signify fragmentation. Although we cannot formally demonstrate this, numerical
  simulations suggest that gravito-turbulence and fragmentations are the
  only outcomes of classic GI. Thus the failure to maintain the 
  gravito-turbulent steady state seems a reasonable description of
  fragmentation. 

Indeed, our linear analysis for perturbations with respect to the 
gravito-turbulent state predicts that PPDs will fragment under similar
conditions observed in numerical simulations \citep[e.g. $\beta_c\sim
  3,\,\alpha_c\sim0.1$,][]{gammie01,rice05}. 

\subsection{Plan}\label{plan}

The basic equations, disk equilibria, cooling and viscosity models are 
given in \S\ref{basic}. The linear stability problem is defined in 
\S\ref{linear}. We present results with parameterized `beta' cooling in 
\S\ref{result_2d} and \S\ref{3ddisk} for 2D and 3D disks,
respectively. 
In \S\ref{2dppd} we consider PPDs with realistic viscosity/cooling   
models, including radiative diffusion in 3D, and determine where PPDs
are gravitationally unstable and why. We summarize our results in
\S\ref{summary} with a discussion of how our models may aid the 
physical understanding of fragmentation in realistic PPDs.   

\input{model3d}

\input{linear3d}
\input{results2d}
\input{results3d}
\input{application}

\input{summary}


We thank the anonymous referee's prompt report that helped to improve
the clarity of this work. We thank S. Stahler and A. Youdin for
comments during the  course of this project; C. Clarke,
C. Gammie, and S.-J. Paardekooper for feedback on the first version of
this article, and G. Lodato for useful discussions.


\appendix
\input{appendix3d}


\end{document}

%% file: model3d.tex
\section{Basic equations}\label{basic}
We consider a 3D, self-gravitating,
viscous disk with heating and cooling. We use the shearing box framework to
study a small patch of the disk \citep{goldreich65}. The local frame
co-rotates with a fiducial point in the unperturbed disk at angular
frequency $\Omega$. The Cartesian co-ordinates $(x,y,z)$ correspond
to the radial, azimuthal and vertical directions in the global disk. 
The fluid equations are
\begin{align}
  &\frac{\p\rho}{\p t} + \nabla\cdot\left(\rho\bm{v}\right) = 0, \\
  & \frac{\p\bm{v}}{\p t} + \bm{v}\cdot\nabla\bm{v} =
  -\frac{1}{\rho}\nabla P - \nabla\Phi - 2\Omega\hat{\bm{z}}\times\bm{v}\notag\\ & 
  \phantom{\frac{\p\bm{v}}{\p t} + \bm{v}\cdot\nabla\bm{v} =}
  +2\Omega^2qx\hat{\bm{x}} -\Omega_z^2z\hat{\bm{z}} 
   + \frac{1}{\rho}\nabla\cdot\bm{T},\label{momentum_eq}\\
  & \frac{\p E}{\p t} + \nabla\cdot\left(E\bm{v}\right) = - P
  \nabla\cdot\bm{v} + \mathcal{H}_\mathrm{visc} - \Lambda +
  \mathcal{H}_\mathrm{ext} ,\label{energy_eq}
\end{align} 
where $\rho$ is the density field and $\bm{v} = (v_x, v_y, v_z)$ is
the velocity field. 
We assume an ideal gas so that the pressure $P$ and thermal energy
density $E$ are related by  
\begin{align}\label{ideal_eos}
  P = (\gamma-1)E = \mathcal{R}\rho T, 
\end{align}
where $\mathcal{R}$ is the gas constant and $T$ is the temperature. 
 For simplicity we refer to the adiabatic index $\gamma$ as that
  in Eq. \ref{ideal_eos} for both 2D and 3D disk models 
  \citep[cf.][]{johnson03}.  The gas 
gravitational potential $\Phi$ is given via the Poisson equation,  
\begin{align}
 \nabla^2\Phi = 4 \pi G \rho. 
\end{align}

In the momentum equation (Eq. \ref{momentum_eq}), the third,
fourth/fifth, and last term on the right-hand side represent the Coriolis, 
tidal, and viscous forces (see below), respectively. We consider    
Keplerian disks with shear parameter $q=3/2$ and vertical oscillation 
frequency $\Omega_z=\Omega$.  

In the energy equation (Eq. \ref{energy_eq}) the source terms 
$\mathcal{H}_\mathrm{visc}$ and $\Lambda$
represent viscous heating and time-dependent cooling, 
respectively, and $\mathcal{H}_\mathrm{ext}$ represents any
time-independent heat source/sinks. We set
$\mathcal{H}_\mathrm{ext}=~0$ unless otherwise stated.  

\subsection{Viscosity and heating}\label{visc_model}
The Cartesian components of the viscous stress tensor $\bm{T}$ are
defined by 
\begin{align}
  T_{ij} \equiv \rho \left[\nu \left(\p_j v_i + \p_i v_j\right) +
    \left(\nu_b-\frac{2}{3}\nu\right)\delta_{ij}\nabla\cdot\bm{v}\right], 
\end{align}
where $\rho\nu$ is the shear viscosity. We also include a bulk
viscosity $\rho\nu_b$ for completeness, but will neglect it in
numerical calculations. The associated viscous heating is given by 
\begin{align}
  \mathcal{H}_\mathrm{visc} \equiv \left(\p_jv_i\right)T_{ij}, 
\end{align}
where summation over repeated indices is implied. 

We adopt a viscosity law  
\begin{align}\label{visc_law}
  \nu = \alpha
  \frac{c_{s0}^2}{\Omega}\left(\frac{\rho}{\rho_\mathrm{eq}}\right)^\mu\left(\frac{P}{P_\mathrm{eq}}\right)^\lambda,             
\end{align}
where subscript `eq' denotes the equilibrium state and  
$c_{s0}^2\equiv P_\mathrm{eq}(z=0)/\rho_\mathrm{eq}(z=0)$. 
The dimensionless viscosity coefficient 
$\alpha=~\alpha(\rho_\mathrm{eq},P_\mathrm{eq})$ characterizes the
magnitude of the shear viscosity \emph{in steady state}. The indices
$\mu,\,\lambda$ are free parameters chosen to model how the viscosity
behaves in the perturbed state.  
We adopt the same prescription for the bulk viscosity but
with $\alpha\to\alpha_b$.   

Our numerical calculations use $\mu=-1,\lambda=0$ so 
that $\rho\nu$ is time-independent, following previous 
studies of viscous GI  
\citep{lynden-bell74,hunter83,willerding92,gammie96}.  
This choice eliminates viscous over-stability
\citep{schmit95,latter06}, which is unrelated to self-gravity, and
would otherwise contaminate our results. 

While steady-state viscosity values can be determined 
analytically or numerically \citep[e.g.][]{martin11,kratter08,rafikov15}, the time-dependent behavior is
not well-explored. We emphasize the choice $\mu=-1,\lambda=0$ is made
to bring out the physical process of interest --- viscous GI. 
Interestingly, though, \cite{laughlin96b} have suggested a $\nu \propto 1/\Sigma$
dependence when modeling the evolution of 2D self-gravitating disks 
as a viscous process. We note that if $\rho\nu$ is constant in time
then increasing the density corresponds to reduction in the viscosity. 
This is perhaps consistent with numerical simulations of disk
fragmentation which show that the internal flow of high-density clumps
is laminar \citep{gammie01}. However, once clumps form they may
effectively decouple from the background disk state, and thus no 
longer be described by the same prescription. 

\subsection{Steady states and cooling models}\label{steady_state}
We consider equilibrium solutions (here omitting the `eq' subscripts
for simplicity)  
\begin{align}
  \bm{v} &= -q\Omega x \hat{\bm{y}}, \\
  \rho   &= \rho(z),\\
  P      &= P(z) \equiv c_s^2(z)\rho.
\end{align} 
The equilibrium density and pressure fields are obtained by solving
the vertical momentum equation with self-gravity,
\begin{align}
  &\frac{1}{\rho}\frac{dP}{dz} +
  \Omega_z^2z + \frac{d\Phi}{dz} = 0, \label{vert_eq1}\\
 &\frac{d^2\Phi}{dz^2} = 4 \pi G \rho,\label{vert_eq2}
\end{align}
together with thermodynamic equilibrium,
\begin{align}\label{thermal_eq}
(q\Omega)^2\rho\nu + \mathcal{H}_\mathrm{ext} = \Lambda,
\end{align}
where the first term represents viscous heating. For the viscous
problem, $\nu\neq0$, and we set $\mathcal{H}_\mathrm{ext}=0$ to 
obtain a relation between viscous heating and cooling 
(e.g. Eq. \ref{alpha_beta_relation} below). However,
if we wish to neglect viscosity (and the accompanying dissipation) but include cooling, we must invoke
$\mathcal{H}_\mathrm{ext}\neq0$ to define an equilibrium state.  
To proceed further, we separately describe the two cooling models
considered in this work.

\subsubsection{Beta cooling}\label{beta_cool_model}
In our beta cooling model, the energy loss per unit volume is specified  
as an explicit function of the thermodynamic variables. 
A prototypical example is 
\begin{align}\label{beta_cool}
  \Lambda(\rho, T) =
  \frac{\mathcal{R}\rho}{(\gamma-1)}\frac{\left(T-T_\mathrm{irr}\right)}{t_c},
\end{align}
 where $T_\mathrm{irr}$ is a reference temperature field, and
recall $\tcool = \beta\Omega^{-1}$ is 
the cooling timescale with $\beta$ a constant input parameter. 
Physically, $\tirr$ may be the floor temperature set by, 
for example, stellar or background irradiation. 

Beta cooling of the form Eq. \ref{beta_cool} is widely applied in 2D
and 3D numerical simulations of self-gravitating disks \citep{gammie01,
  rice05,rice11,paardekooper12}. In fact, for $T_\mathrm{irr}=0$ the 
cooling function $\Lambda = E/t_\mathrm{c}$ is identical 
to that originally employed by \cite{gammie01}. 
We will refer to Eq. \ref{beta_cool} as  
`standard' beta cooling. It permits numerical experiments to be
carried out in a controlled manner as a function of the cooling time
$\beta$. An adiabatic disk corresponds to $\beta\to \infty$. The
physical meaning of the limit $\beta\to0$ depends on $T_\mathrm{irr}$, as
discussed in \S\ref{theta0}---\ref{theta1}.  

 For standard beta cooling we assume an 
equilibrium polytropic relation 
\begin{align}\label{poly_vert} 
  P  =
c_{s0}^2\rho_0\left(\frac{\rho}{\rho_0}\right)^\Gamma,
\end{align}
 where $\rho_0 = \rho(z=0)$ is the equilibrium mid-plane density, and    
$\Gamma$ is the constant polytropic index that determines the disk's
  vertical structure. Thus $\Gamma$ is only relevant to the 3D
  problem.    
The vertical structure is first obtained from 
Eq. \ref{vert_eq1}---\ref{vert_eq2}, 
then inserted into 
Eq. \ref{thermal_eq} to infer the required viscosity profile for
thermal equilibrium. If  $\mathcal{H}_\mathrm{ext}=0$, 
\begin{align}\label{alpha_beta_relation}
\alpha(z) = \frac{1}{(\gamma-1)\beta
   q^2}\frac{c_s^2(z)}{c_{s0}^2}\left(1 - \theta\right),
\end{align} 
 with 
  \begin{align}\label{tirr_def}
    \theta = \frac{\tirr}{T_\mathrm{eq}},
  \end{align}
   where $T_\mathrm{eq}(z)$ is the equilibrium temperature field.
   We shall consider vertically isothermal disks with $\Gamma=1$, as
  appropriate for the outer parts of irradiated protoplanetary disks
   \citep{chiang97}. In this case  
   $\alpha$, $T_\mathrm{eq}$ and $\theta$ are simply constants. 
    Note that $\theta$  
    should only be interpreted as an irradiation parameter when it is 
    defined through Eq. \ref{tirr_def}, in conjunction with adopting  
    Eq. \ref{beta_cool} as the cooling function. 


We assume standard beta cooling in formulating the linear
problem. However, the corresponding linear problem for  
any other explicit cooling function, say $\Lambda_1(\rho,T)$, can be  
obtained by equating its linearized form to 
that of Eq. \ref{beta_cool}, i.e. setting $\delta \Lambda \equiv
\delta \Lambda_1$. This then defines the $\beta$ and $\theta$
parameters to be used in the framework we develop later (see also 
\S\ref{linear_bcool}). We do this in \S\ref{2dppd} where we adopt a  
more realistic beta cooling function for PPDs. In that case, $\theta$ 
may or may not directly represent a physical irradiation.     

\subsubsection{Radiative cooling}\label{rad_cool}
A more realistic treatment of cooling considers energy transfer by radiative diffusion. Then
\begin{align}
  \Lambda &= \nabla\cdot\bm{F}_\mathrm{rad},\label{rad_cool1}\\
  \bm{F}_\mathrm{rad}   &= -\frac{16\sigma T^3}{3\kappa_d\rho}\nabla T, \label{rad_cool2}
\end{align}
where $\sigma$ is the Stefan-Boltzmann constant and 
$\kappa_d$ is the (dust) opacity. We adopt
\begin{align}\label{opacity_law}
  \kappa_d = \kappa_{d0}T^b,
\end{align}
and take the constant index $b=2$ as appropriate for the cold outer regions
of a PPD with ISM-like dust grains \citep{bell94}, but retain the general notation $b$
to keep track of the opacity.    

In this case, we specify a constant viscosity coefficient $\alpha$ and
solve Eq. \ref{vert_eq1}---\ref{thermal_eq}, together with
Eq. \ref{rad_cool1}---\ref{rad_cool2}, as a fourth order system of
ordinary differential equations to obtain equilibrium profiles $P(z)$, $T(z)$, and hence
$\rho(z)$.  

While radiative cooling is arguably more realistic than
beta cooling, it generally implies a vertically
non-isothermal equilibrium disk, and increases the order of the
linearized equations. It formally applies to optically-thick 
disks, but it is possible to modify the flux function to account
for optically-thin disks \citep{levermore81}. However, this
complication is beyond the scope of this work.

%% file: linear3d.tex
\section{Linear problem}\label{linear}
We consider infinitesimal axisymmetric Eulerian perturbations of the
form \begin{align}
  \delta \rho= \widetilde{\delta \rho}(z)\exp{\left(\ii kx +
    s t\right)},  
\end{align}
and equivalent form for other variables. Here, $k$  is an input real horizontal
wavenumber and $s$ is a (generally) complex growth rate.  
For simplicity, hereafter we drop the tilde.

The linearized continuity, momentum and energy equations are
\begin{align}
   &s \delta \rho = - \ii k \rho\delta v_x -
  \left(\rho\delta v_z\right)^\prime \label{lin_mass} \\ 
   &s  \delta v_x = - \ii k\frac{\delta P}{\rho} - \ii k \delta 
  \Phi + 2\Omega \delta v_y + \delta F_x,\\
   &s \delta v_y = (q-2)\Omega\delta v_x + \delta F_y,\\
  &s \delta v_z =
  \left(\ln{P}\right)^\prime c_s^2\frac{\delta\rho}{\rho} -
  \left(\frac{\delta P}{\rho}\right)^\prime -
  \left(\ln{\rho}\right)^\prime\frac{\delta P}{\rho} + \delta F_z, \\ 
  &s \frac{\delta P}{\rho} = 
 - \ii k \gamma c_s^2\delta v_x
- c_s^2\left[
 \left(\ln{P}\right)^\prime\delta v_z + 
  \gamma\delta v_z^\prime\right] \notag\\
   & \phantom{s \frac{\delta P}{\rho}=}
  + (\gamma-1)\frac{\delta\mathcal{H}_\mathrm{visc}}{\rho} 
  -(\gamma-1)\frac{\delta\Lambda}{\rho}
  \label{lin_energy},\\
  &\delta\Phi^{\prime\prime} - k^2\delta\Phi =
  \frac{\Omega^2}{Q_\mathrm{3D}}\left(\frac{\delta\rho}{\rho_0}\right)\label{lin_gravity},  
\end{align}
where $^\prime$ denotes $d/dz$. The perturbed viscous forces are 
\begin{align}
  \delta F_x =& \nu \left[\delta v_x^{\prime\prime} + 
    \left(\ln{\rho\nu}\right)^\prime \delta v_x^\prime - 
     \frac{4}{3}k^2\delta v_x \right] - \nu_bk^2\delta v_x \notag\\
   &+ \ii \nu k \left[\frac{1}{3}\delta v_z^\prime+ 
    \left(\ln{\rho\nu}\right)^\prime\delta v_z\right] + \ii k \nu_b
  \delta v_z^\prime,\\
  \delta F_y = & \nu  \left[\delta v_y^{\prime\prime} + 
    \left(\ln{\rho\nu}\right)^\prime \delta v_y^\prime - 
     k^2\delta v_y \right] - \ii\nu kq\Omega\delta\ln{\rho\nu},\\
  \delta F_z =& \nu\left[\frac{4}{3}\delta v_z^{\prime\prime} + 
    \frac{4}{3}\left(\ln{\rho\nu}\right)^\prime \delta v_z^\prime - 
    k^2\delta v_z \right] \notag\\
  & +\nu_b\left[\delta v_z^{\prime\prime} + \left(\ln{\rho\nu_b}\right)^\prime\delta v_z^\prime\right]\notag\\
  & + \ii \nu k  \left[\frac{1}{3}\delta v_x^\prime -  
    \frac{2}{3}\left(\ln{\rho\nu}\right)^\prime\delta v_x\right]\notag\\
  &+\ii \nu_b k \left[\delta v_x^\prime + \left(\ln{\rho\nu_b}\right)^\prime\delta v_x\right],
\end{align}
and the perturbed viscous heating is given by 
\begin{align}
  \frac{\delta\mathcal{H}_\mathrm{visc}}{\rho}=& \nu (q\Omega)^2
  \delta\ln{\rho\nu} - 2\ii\nu k q \Omega \delta v_y,\\
  \delta\ln{\rho\nu} =& (1+\mu)\frac{\delta\rho}{\rho} +
  \frac{\lambda}{c_s^2}\frac{\delta P}{\rho}. \label{linear_beta_cool}
\end{align}
In Eq. \ref{lin_gravity}, the 3D self-gravity parameter is
\begin{align}\label{Q3d_def}
  Q_\mathrm{3D} \equiv \frac{\Omega^2}{4\pi G \rho_0} 
\end{align}
\citep{mamat10}. 
The linearized cooling functions $\delta\Lambda$ are given below. 
Eq. \ref{lin_mass}---\ref{lin_gravity}, supplemented with appropriate
boundary conditions, constitutes an eigenvalue problem for the growth
rate $s$. 

\subsection{Linearized beta cooling}\label{linear_bcool}
For the standard beta cooling prescription, linearizing
Eq. \ref{beta_cool} gives

\begin{align}
  (\gamma-1)\frac{\delta \Lambda}{\rho} = \frac{1}{t_c}\left(\frac{\delta P}{\rho} -
  \theta c_s^2 \frac{\delta\rho}{\rho}\right), \label{linear_beta}
\end{align}
where we have used $\delta T/T  = \delta P/P - \delta\rho/\rho$ from
the ideal gas law. 

Note that any beta cooling function can be linearized in the
form of Eq. \ref{linear_beta} with appropriate definitions of $t_c$
and $\theta$ (see \S\ref{ppd_cooling} for an example).    
For the stability problem we may simply regard $\theta$ as a parameter 
for the density-dependence of any generic beta cooling function. If 
we specifically consider standard beta cooling, then $\theta$ also
represents physical irradiation.     

\subsection{Linearized radiative cooling}
Linearizing Eq. \ref{rad_cool1}---\ref{rad_cool2} with the
temperature-dependent opacity law in Eq. \ref{opacity_law} gives
\begin{align}
  \frac{\delta \Lambda}{\rho} 
  =& \frac{16\sigma T^3}{3\kappa_d\rho^2} k^2 \delta T \notag\\
  & - \frac{16\sigma }{3\rho}\frac{d}{dz}\left\{
  \frac{T^3}{\kappa_d\rho}\left[\delta T^\prime +
    (3-b)\left(\ln{T}\right)^\prime \delta T - T^\prime \frac{\delta\rho}{\rho}\right]
  \right\}.\label{linear_rad_cool}
\end{align}
Since Eq. \ref{linear_rad_cool} contains vertical derivatives of the
perturbations, it is not generically possible to map radiative cooling 
to the beta cooling prescription, except for special problems
\citep[e.g.][]{lin15}. 

We now consider the gravitational stability of two and three 
dimensional disks in the presence of non-ideal
physics: cooling and viscosity.  

%% file: results2d.tex
\section{Two-dimensional disks with beta cooling}\label{result_2d}
We begin in the 2D limit with standard beta cooling  to facilitate comparison 
with previous studies. The disk
material is assumed to be confined to the mid-plane, and $\delta v_z=0$. 
We make the replacement  
$\rho \to \Sigma$, re-interpret $P$ as the vertically-integrated
pressure, and set $\Gamma=1$. 
The gravitational potential perturbation remains 3D and its mid-plane
value is given by    
\begin{align}
  \delta \Phi(z=0) = -\frac{2 \pi G}{|k|}\delta\Sigma
\end{align}
\citep{shu70}.
 
The linearized equations yield an algebraic dispersion relation $s =
s(k)$. We write this  
in terms of the dimensionless growth rate $S = s/\Omega$ and
wavenumber $K=kH = k c_{s0}/\Omega$ as
\begin{align}\label{thindisk}
  f(S,K)\equiv AD - BC = 0,   
\end{align}
where the functions $A,B,C,D$ are given in Appendix \ref{2ddisp}. 
We use this generalized dispersion relation  to investigate GI driven 
by cooling in \S\ref{2d_inviscid}; and GI driven by viscosity 
in \S\ref{2dvisc}.

\subsection{Inviscid limit}\label{2d_inviscid}
We first simplify the problem by setting $\alpha = \alpha_b = 0$. This
eliminates viscous heating and forces in the linearized problem, 
allowing us to quantify the \emph{sole} effect of cooling on the 
perturbations. 
We emphasize that destabilization is independent of the effect of
decreasing temperature on the instantaneous value of the
classic-$Q$. A time-independent heat source should be invoked to
balance the imposed cooling to allow an equilibrium to be 
defined ($\mathcal{H}_\mathrm{ext}\neq 0$).

For example, we could assume that the viscosity only provides a
background heating and does not play an active role in the perturbed
state. This is in fact done implicitly in the literature when
discussing fragmentation of cooling, 
self-gravitating disk simulations \citep{gammie01}. 
There, the effect of the ambient 
gravito-turbulent viscosity on the forming-clump is neglected, as 
one only compares adiabatic heating and the imposed cooling.  
This comparison is encapsulated in the generalized dispersion relation
below.


Eq. \ref{thindisk} becomes 
\begin{align}\label{inviscid}
  S^2 = \frac{2|K|}{Q} - 2(2-q) - \left(\frac{\theta + \beta \gamma
    S}{1+\beta S}\right)K^2, 
\end{align}
similar to the classic dispersion relation
(Eq. \ref{classic_dr}), which may be obtained by taking the limit
$|\beta S|\to\infty$. 
The first term on the right-hand-side represents destabilization by self-gravity; 
the second and third terms represent stabilization by rotation and
pressure, respectively. The imposed cooling/irradiation only affects the
pressure response. 

Eq. \ref{inviscid} is a cubic equation in $S$. The 
Routh-Hurwitz criteria imply that stability is ensured if 
\begin{align}\label{stable_condition}
  \gamma > \theta \quad \text{and} \quad 
  Q > \frac{1}{\sqrt{2\theta(2-q)}} 
\end{align}
are both satisfied. 
A third criterion, $2(2-q)\gamma Q^2>1$, is formally required, but this
is implied by Eq. \ref{stable_condition}. 
Notice these conditions do not actually depend on the cooling time. 
At fixed $Q$, the second stability condition eventually fails for
decreasing $\theta$, i.e. if perturbations are allowed
to cool to sufficiently low temperatures.  


If only real growth rates are considered, then violating the second
condition in Eq. \ref{stable_condition} alone is sufficient for instability. 
In that case the wavenumbers satisfying
\begin{align}
  \theta K^2 - \frac{2}{Q}|K| + 2(2-q) < 0 
\end{align}
are unstable. The range of unstable wavenumbers
increases with decreasing irradiation $\theta$. For $\theta\ll 1$ this range is 
$(2-q)Q\lesssim|K|\lesssim2/\theta Q$. Without irradiation there is no
upper limit to unstable wavenumbers, which could have implications for
numerical simulations probing large wavenumbers and small scales 
at high resolution (see \S\ref{prev_works}). 

Consider the most unstable wavenumber $|K_*|$ at which $\p S/\p|K| =
0$ and $S = S_*$. By differentiating Eq. \ref{inviscid}, we obtain 
\begin{align}\label{kstar}
  |K_*| = \frac{1+\beta S_*}{Q\left(\theta + \gamma \beta S_*\right)}
\end{align}
Inserting this into Eq. \ref{inviscid}, we find the maximum growth
rate satisfies
\begin{align}\label{inviscid_max}
  S_*^2 = \frac{1+\beta S_*}{Q^2\left(\theta + \beta\gamma S_*\right)}
  - 2(2-q).
\end{align}
Eq. \ref{kstar}---\ref{inviscid_max} imply $\p_\beta|K_*|,\,\p_\beta
S_*<0 $ for $\gamma>\theta$. Then as cooling becomes more rapid, the maximum 
growth rate increases, and the most unstable wavelength decreases. 
For $|S_*|\ll1$, Eq. \ref{inviscid_max} gives the simple solution 
\begin{align}
  \beta S_*\simeq \frac{1-2(2-q)Q^2\theta}{2(2-q)Q^2\gamma-1}.  
\end{align} 
Thus $S_*\to0$ as $\beta\to\infty$, but growth rates are never zero
for any finite $\beta$. That is, the disk can be formally unstable for
arbitrarily long cooling times.       

\begin{figure}
  \includegraphics[width=\linewidth,clip=true,trim=0cm 0cm 0cm
    0.0cm]{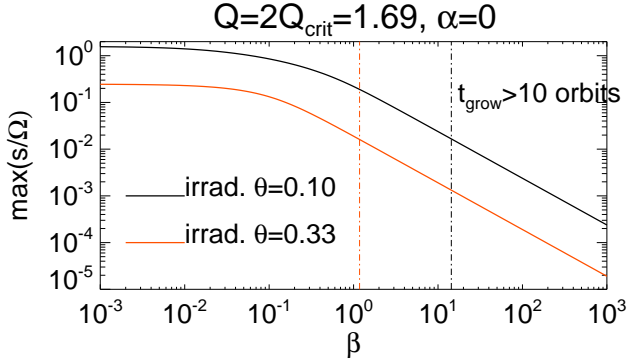}
  \caption{
    Growth rates for the 2D inviscid problem as a function of
    the cooling time $\beta$ for two irradiation levels: $\theta=0.1$ (black) and
    $\theta=0.33$ (orange). For each case the vertical dashed-dotted
    lines mark the cooling times beyond which 
    growth rates are longer than 10 orbits. 
    \label{invisc_theta}}
\end{figure}

Fig. \ref{invisc_theta} shows growth rates in a $Q=1.7$ disk  
as a function of the cooling time $\beta$ for two irradiation levels
$\theta=0.1,\,0.33$. The vertical lines mark characteristic cooling
times beyond which the growth timescale is long compared to the
dynamical time. Increasing irradiation stabilizes the
disk, and faster cooling is required to achieve the same growth 
rate as in a disk with weaker irradiation.  

Next, we consider the two limiting cases: $\theta=0$, so perturbations are
cooled towards zero temperature \citep[typically employed in
numerical simulations, e.g.][]{gammie01}; and $\theta=1$, where the
equilibrium disk temperature equals the irradiation temperature.

\subsubsection{$\theta = 0$}\label{theta0}
For $\beta$-cooling with $\theta=0$, the disk is unconditionally
unstable for finite $Q$, although instability occurs on smaller scales
as $Q$ increases.  The $\beta\to0$ limit corresponds to a 
pressureless disk (not merely isothermal).   

Let us consider a disk with 
\begin{align}\label{qcrit_def}
  \qtwo = \frac{1}{\sqrt{2\gamma(2-q)}}\equiv Q_{\mathrm{crit}},
\end{align} 
which is the condition for marginal stability in an adiabatic disk. 
How does finite cooling destabilize the disk?  
Inserting Eq. \ref{qcrit_def} into Eq. \ref{inviscid_max} with 
$\theta=0$, we find 
\begin{align}\label{sstar}
  S_*^3 = \frac{1}{\gamma Q_\mathrm{crit}^2 \beta}. 
\end{align}
The maximum growth rate, $S_*\propto \beta^{-1/3}$, smoothly
increases with decreasing $\beta$. Notice for $\beta>1$ this growth
  rate is faster than the imposed cooling rate $\beta^{-1}\Omega$.

We can define a characteristic cooling
time $\beta_*$ as that which removes pressure support against
self-gravity over the natural lengthscale in the problem, the
scale-height $H$. We thus set $|K_*|=1$ and find, for the Keplerian
disk,  

\begin{align}\label{betastar}
  \beta_* = \frac{1}{\left(\sqrt{\gamma} - 1\right)^{3/2}}. 
\end{align}
This equation gives similar values 
of the cooling times below which 
numerical simulations show dynamical disk fragmentation 
\citep{gammie01,rice05,rice11}. These simulations employ the same beta  
cooling prescription with $\theta=0$, and determine the fragmentation
boundary, $\beta_c$, as a function of the adiabatic index $\gamma$.  
Table \ref{bstar_compare} shows rough agreement between $\beta_*$ and 
$\beta_c$. The match is remarkable, especially with the
  global 3D simulations of \cite{rice05}, since Eq. \ref{betastar}
  is derived for 2D disks in the local limit. 

\begin{deluxetable}{rrrr}
  \tablecolumns{4}
  \tablecaption{Characteristic cooling times as a function of
    $\gamma$. \label{bstar_compare}
  }
  \tablehead{
    \colhead{$\gamma$}   & \colhead{Eq. \ref{betastar}, $\beta_*$} &
    \colhead{Simulation, $\beta_c$} &  \colhead{Reference}
  }
\startdata
 $7/5$ & 12.75 & 12---13 & \cite{rice05}\\
$1.6$  &  7.33 & 8 & \cite{rice11}\\
$5/3$  &  6.37 & 6---7 & \cite{rice05}\\
$2$    &  3.75 & 3 & \cite{gammie01}
\enddata
\end{deluxetable}

\subsubsection{$\theta = 1$}\label{theta1}
Standard beta cooling with $\theta=1$ corresponds to 
`thermal relaxation': the temperature is restored to its initial
value over the cooling time \citep{lin15,mohandas15}. In this case no
additional heat source need be invoked to define an inviscid steady state. From
Eq. \ref{stable_condition}, the instability condition is $Q<1$. 
This is the same as the classic Toomre condition for an isothermal
disk (which may be obtained from Eq. \ref{inviscid}  by taking
$|\beta S|\to 0$). In this respect, a 
fully irradiated disk, in which the equilibrium temperature is set
externally, behaves isothermally regardless of the cooling time 
\citep{gammie01,johnson03}.



%

\subsection{Viscous disk}\label{2dvisc}
We now consider a viscous disk with parameters $\mu=-1,\,\lambda=0$ in our
adopted viscosity law, Eq. \ref{visc_law}. In the 2D case this implies
$\nu\Sigma$ is constant.   
We check in Appendix \ref{gammie_check} that our dispersion relation
reduces to previous results for viscous GI in the isothermal limit (by
taking $|\beta S|\to~\infty$ and $\gamma=1$).   

It is useful to consider several limiting cases.
To see the effect of cooling and irradiation, we simplify the
dispersion relation, Eq. \ref{thindisk}, by assuming $|\beta S|\ll
1$. Then for $|K| \to 0$ we find
\begin{align}\label{gammie_smallk}
  S\simeq \frac{\alpha K^2}{2(2-q)}\left(\frac{2|K|}{Q} - \theta
  K^2\right), 
\end{align}
which coincides with \citeauthor{gammie96}'s Eq. 18 for vanishing
wavenumber. For $|K|\to\infty$ we find
\begin{align}\label{gammie_bigk}
  S \simeq\left(\frac{2}{Q|K|} - \theta\right)\left(\frac{4}{3}\alpha + 
  \alpha_b + \gamma\beta\right)^{-1}.
\end{align}

\begin{figure}
  \includegraphics[width=\linewidth,clip=true,trim=0cm 2cm 0cm
    0.0cm]{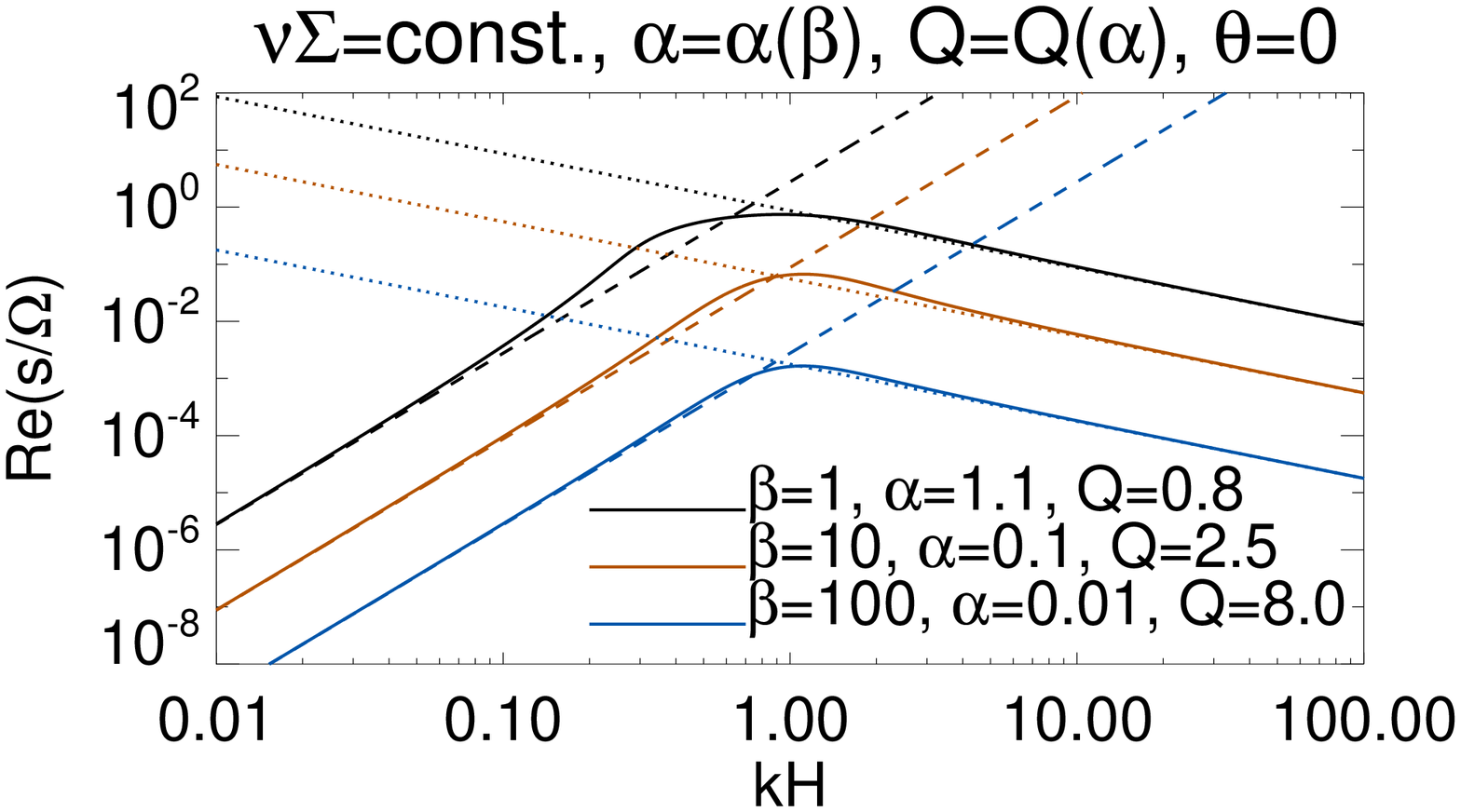}\\
  \includegraphics[width=\linewidth,clip=true,trim=0cm 0cm 0.cm
    0.0cm]{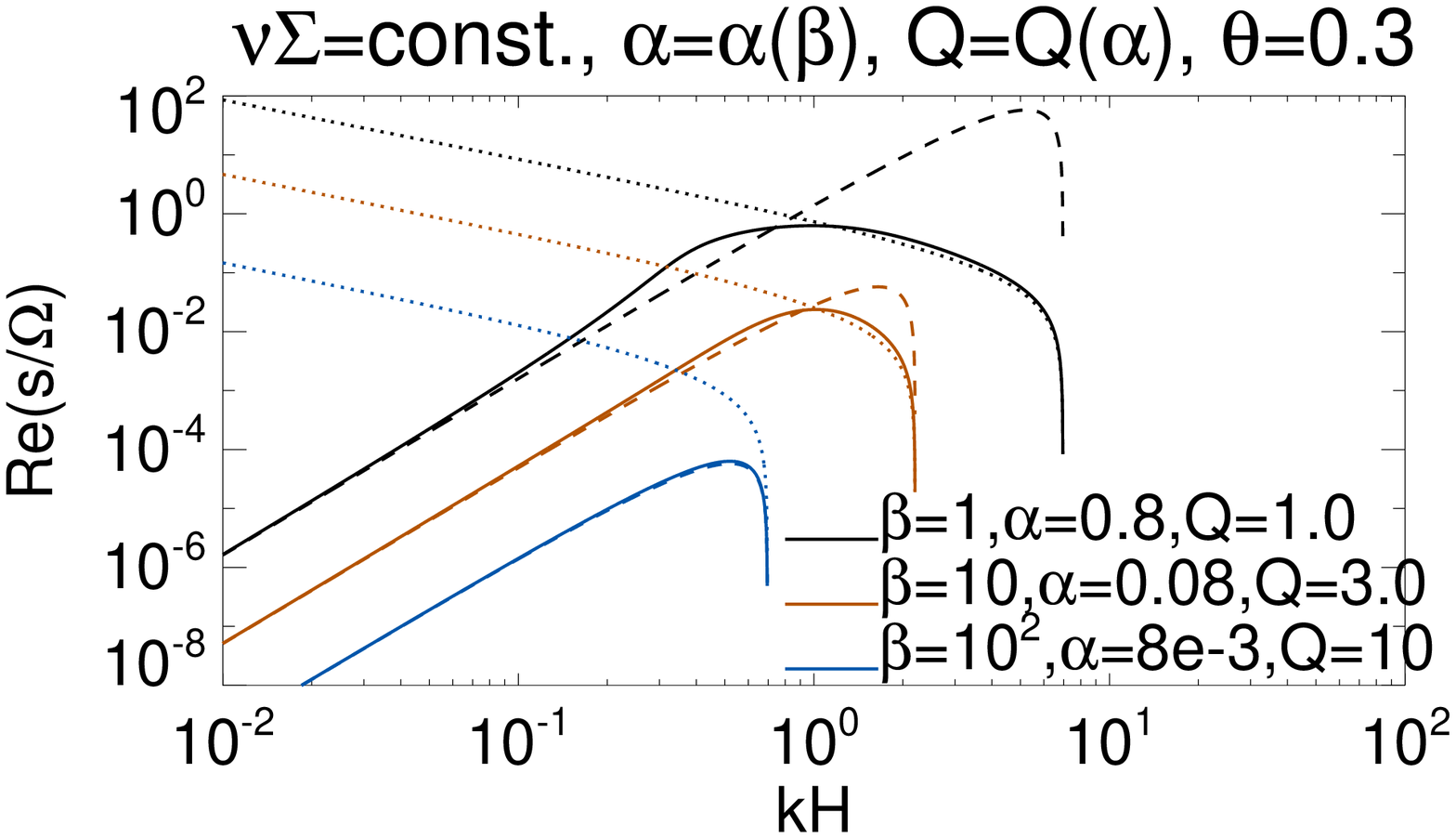}
  \caption{Growth rates for the 2D viscous problem as a function of
    the radial wavenumber, $k$, for a range of cooling times, 
    $\beta$. The dashed and dotted lines correspond to asymptotic
    behaviors for small and large $k$, respectively, computed from
    Eq. \ref{gammie_smallk} and \ref{gammie_bigk}. Top: without
    a floor temperature ($\tirr = 0$); bottom: with a floor
    temperature $\tirr$ set to $30\%$ of the equilibrium temperature.   
    \label{gammie_rate_plot}}
\end{figure}

For $\theta\ll1$ a rough measure of the maximum growth rate can be obtained by
equating Eq. \ref{gammie_smallk} and \ref{gammie_bigk}\footnote{If
  $\theta$ is not small and/or $Q$  is large then one may just use Eq. \ref{gammie_smallk}
  to maximize $S$ over $K$, see the $\theta=0.3,\beta=100$ curve in the bottom
  panel of Fig. \ref{gammie_rate_plot}}.  
This exercise yields 
\begin{align}\label{gammie_maxrate_simple}
  S_*\simeq \frac{
    6^{3/4}\left[\alpha\left(4\alpha +
      3\alpha_b + 3\gamma\beta\right)\right]^{1/4} - 3\theta
    Q(2-q)^{1/4}}{Q\left(4\alpha + 3\alpha_b +
    3\gamma\beta\right)(2-q)^{1/4}}. 
\end{align} 

To compute growth rates numerically, 
we consider a model with $\gamma=1.4$, $\alpha_b=0$ and 
$\alpha=\alpha(\beta)$ given by thermal equilibrium
(Eq. \ref{alpha_beta_relation}). Furthermore, we relate
 the strength of self-gravity and viscosity by   
\begin{align} 
  Q = \frac{Q_\mathrm{crit}}{\sqrt{\alpha}},\label{Qalpha}
\end{align}
to mimic a gravito-turbulent basic state, where one
might expect the dimensionless stress $\alpha \sim Q^{-2}$
\citep{lin87}.   



Fig. \ref{gammie_rate_plot} shows growth rates as a function of the
wavenumber obtained from the dispersion relation 
Eq. \ref{thindisk}. The limiting behavior for small/large $K$ are
well-captured by Eqs. \ref{gammie_smallk} and
\ref{gammie_bigk}. Comparing the two panels 
shows that increasing the
irradiation level ($\theta$) suppresses small-scale
perturbations. 


\begin{figure}
  \includegraphics[width=\linewidth,clip=true,trim=0cm 2cm 0.cm
    0.0cm]{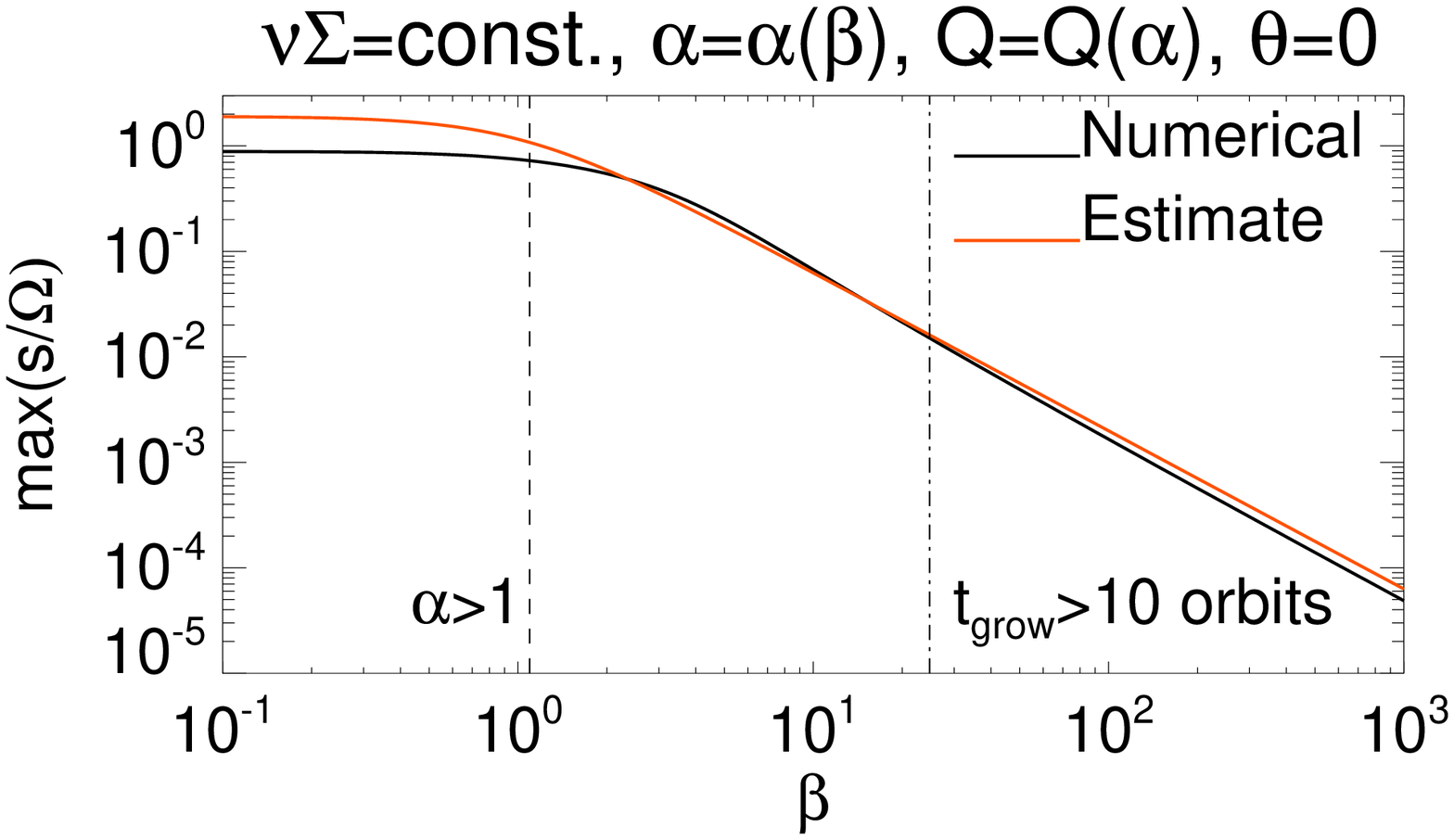}
  \includegraphics[width=\linewidth,clip=true,trim=0cm 0cm 0.cm
    1.0cm]{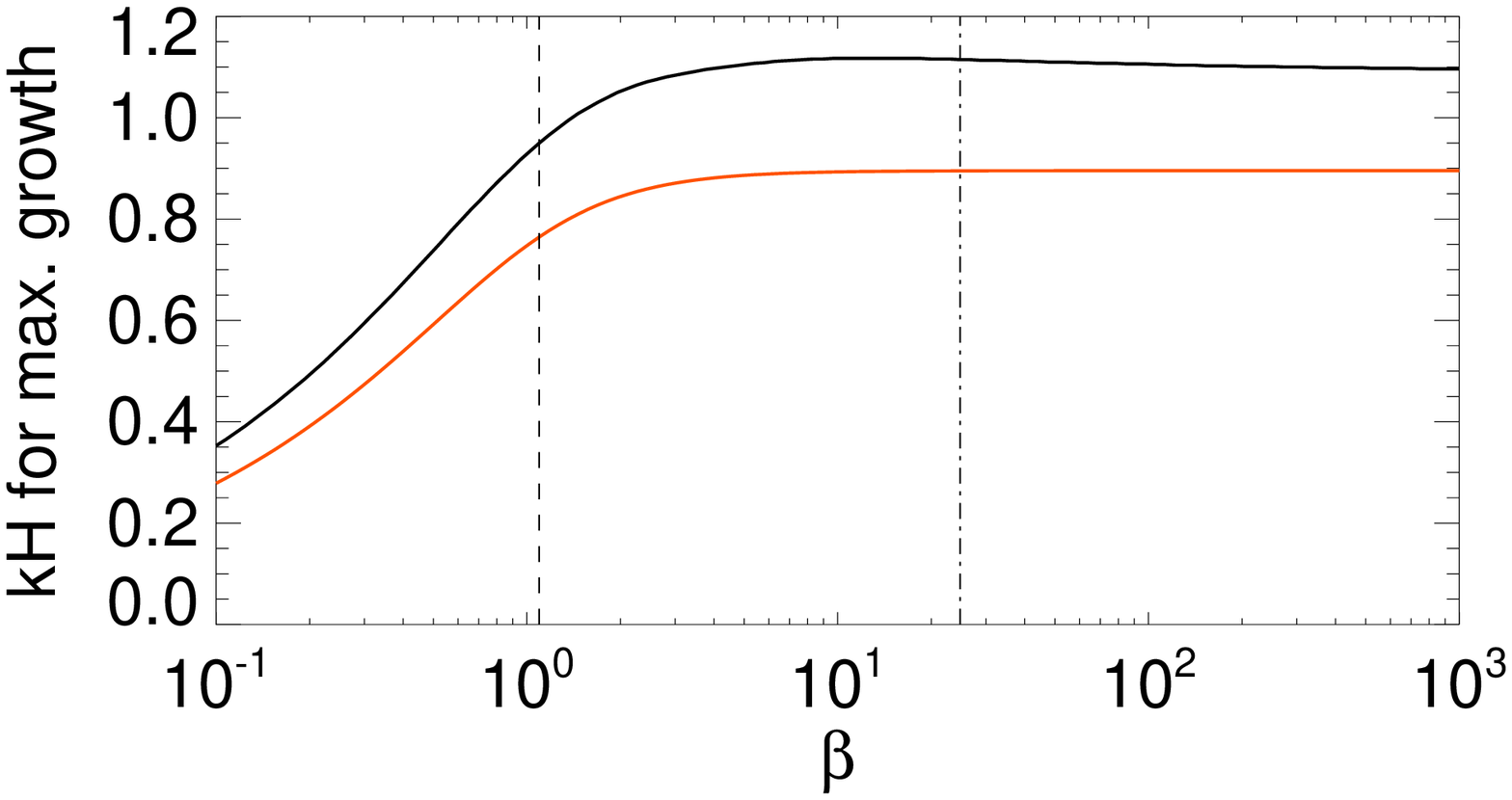}
  \caption{Growth rates (top) at the optimal wavenumber (bottom) for
    viscous GI as a function of cooling time for the
    case shown in the top panel of Fig. \ref{gammie_rate_plot}. Black curves 
    computed from the dispersion relation (Eq. \ref{thindisk}), and
    the orange curves are estimates based on
    Eq. \ref{gammie_smallk}---\ref{gammie_bigk}. 
    The dashed
    line marks the region with $\alpha > 1$, and the 
    dashed-dot line marks the region where growth timescales exceed
    the dynamical time.   
    \label{gammie_maxrate_plot}}
\end{figure}

Fig. \ref{gammie_maxrate_plot} shows the maximum
growth rate (top panel) and the corresponding wavenumber (bottom
panel) as a function of the cooling time $\beta$ for $\theta=0$.  
There is good agreement between numerical growth rates and
Eq. \ref{gammie_maxrate_simple} for $\beta \gtrsim 1$. 
Eq. \ref{gammie_maxrate_simple} gives the
limiting behavior for this case as   
\begin{align}
  S_*\propto \begin{cases}
    \alpha^{1/4}\beta^{-3/4}Q^{-1}\propto \beta^{-3/2} 
    &  \beta \gg \alpha, \\
    \alpha^{-1/2}Q^{-1} = \mathrm{const.} & \beta \ll \alpha, \label{high_visc}   
  \end{cases}
\end{align}
where we have applied Eq. \ref{alpha_beta_relation}
and \ref{Qalpha}. 
The disk is unstable for all $\beta$, but
growth timescales are long ($>10$ orbits) for $\beta \gtrsim
20$. 
This region is marked by the vertical dashed-dotted line in
Fig. \ref{gammie_maxrate_plot}. The optimum wavenumber decreases with
the cooling time for $\beta\lesssim O(1)$ because larger scales are 
more resistant to the associated increase in viscous damping. This is
evident from the dispersion relation in the large wavenumber limit,
Eq. \ref{gammie_bigk}, showing increasing viscosity weakens
small-scale modes. 

Numerical simulations of gravito-turbulent disks show there is a
maximum $\alpha$ ($\sim 0.06$) that can be sustained before
fragmentation \citep{rice05}. We can interpret this result in our
linear framework. 
Suppose it is possible to balance rapid cooling ($\beta\lesssim1$) 
by generating a large gravito-turbulent heating rate
($\alpha\gtrsim 1$) through a small $Q$ (second case in
Eq. \ref{high_visc}). Fig. \ref{gammie_maxrate_plot}  
shows that such a disk would be dynamically unstable with 
growth rate $s = O(\Omega)$. This is due to the direct effect of viscous 
stress promoting instability, rather than cooling. Thus, we do not expect rapidly-cooled, 
and hence highly turbulent, self-gravitating disks to persist beyond
dynamical timescales.
We might interpret fragmentation as an instability
of the highly viscous state (see \S\ref{gi_of_gi}). This is
consistent with previous numerical 
simulations performed by \cite{lodato05}. 

%% file: results3d.tex
\section{Three-dimensional disks with beta cooling}\label{3ddisk}
We confirm the above results in 3D disks with vertical
structure. Accounting for the third dimension will weaken gravitational instabilities
because the disk mass is spread across some vertical extent. It is 
possible to incorporate this effect in the previous 2D framework, but doing so
introduces an additional `softening' parameter $H_\mathrm{sg}$ as discussed in
Appendix \ref{3dcorr}. It is more direct to solve the 3D eigenvalue
problem to avoid such uncertainties. Our numerical approach is
outlined in Appendix \ref{3d_method}. 


In the following examples we consider a vertically isothermal disk
($\Gamma=1$ in Eq. \ref{poly_vert}). Then in the viscous case $\alpha$ is vertically
constant (Eq. \ref{alpha_beta_relation}). We consider only even modes
about the mid-plane, and apply a numerical disk surface at $z=\zmax$
such that $\rho(\zmax)=0.05\rho_0$.     

\subsection{Inviscid 3D disk}

We consider a 3D inviscid disk with $\gamma=1.4$, $\qthree=0.71$, and $\theta=0$. 
The 3D gravity parameter $\qthree$ is defined by 
Eq. \ref{Q3d_def}. For such a disk, the corresponding Toomre parameter 
$Q=2Q_\mathrm{crit}$. Recall $Q_\mathrm{crit}$, defined by
Eq. \ref{qcrit_def}, is the Toomre parameter value such that the 2D
disk would be marginally stable in the absence of cooling. 

\begin{figure}
  \includegraphics[width=\linewidth,clip=true,trim=0cm 2cm 0.cm
    0.0cm]{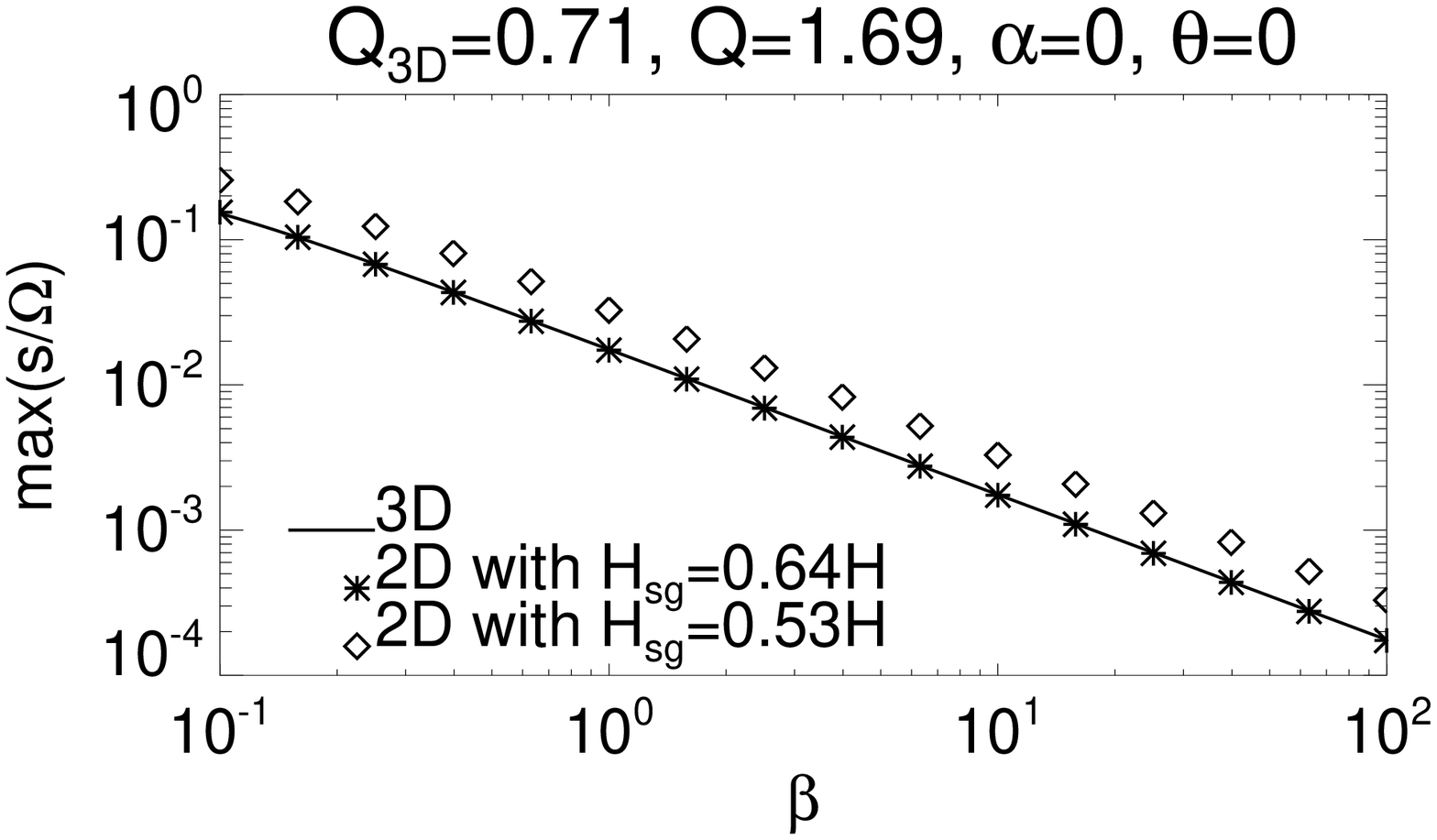}\\
  \includegraphics[width=\linewidth,clip=true,trim=0cm 0cm 0.cm
    0.8cm]{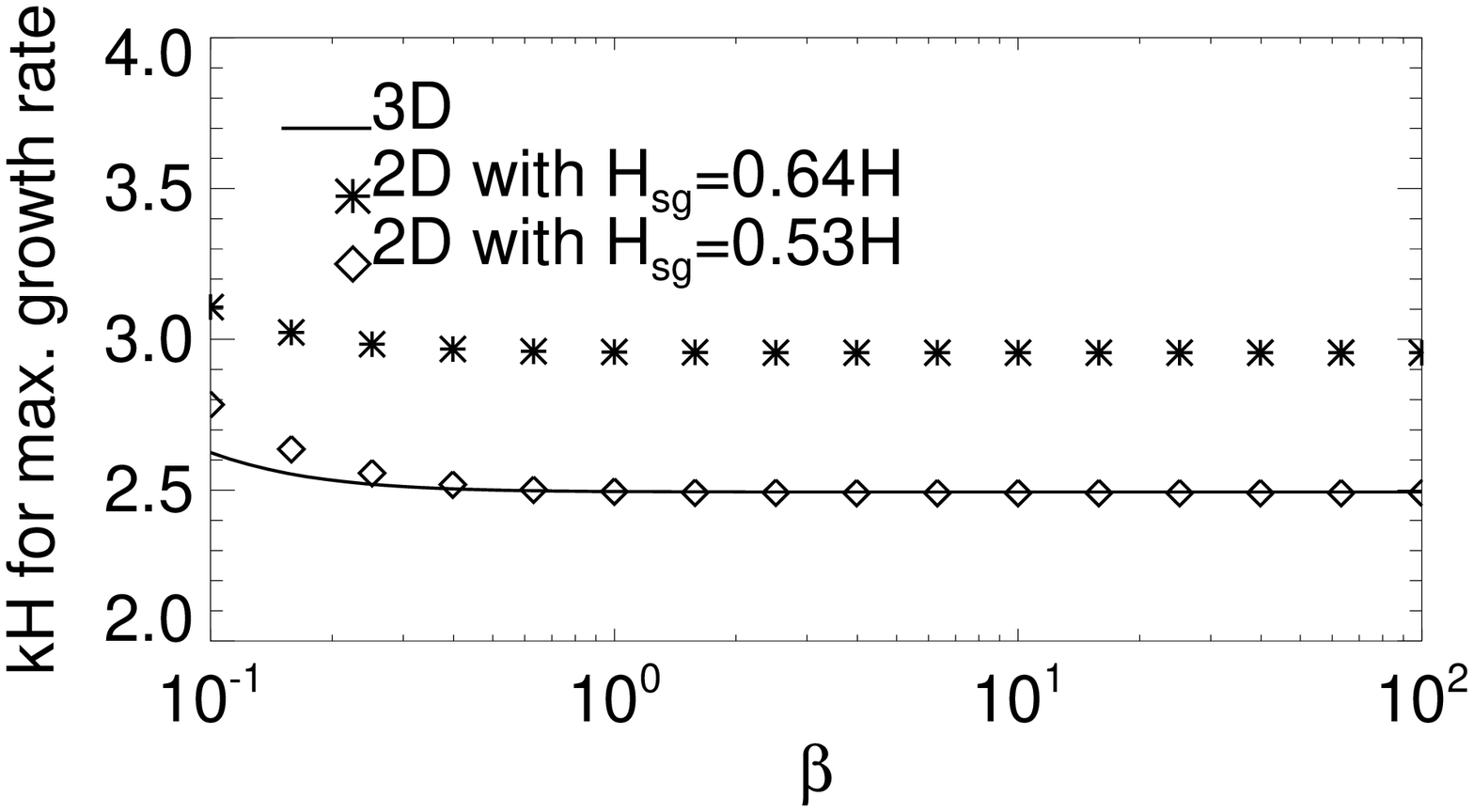}
  \caption{Growth rates (top) and optimal wavenumber (bottom) obtained
    from the inviscid 3D eigenvalue problem (solid line). Asterisks
    and diamonds are corresponding values from the 2D dispersion
    relation (Eq. \ref{thindisk}) but with a softened gravity as
    described in Appendix \ref{3dcorr}. \label{3d_inviscid}} 
\end{figure}

Fig. \ref{3d_inviscid} shows growth rates and the most unstable
wavenumbers obtained for this model. We also plot 2D
results with the 3D correction as described in Appendix
\ref{3dcorr}. The (empirically) chosen value of 
$H_\mathrm{sg}=0.64H$ results in a close match between 2D and 3D
growth rates, but the most unstable wavenumber in 3D is somewhat smaller. 
This offset reflects self-gravity being weakened in the vertical 
direction: a larger horizontal scale is required to achieve the same
strength of self-gravity as the 2D case. Similarly, 
choosing $H_\mathrm{sg}=0.53H$ matches the optimum wavenumbers, but
growth rates are over-estimated in 2D.    

\subsection{Viscous 3D disk} 
For the 3D viscous problem we use the same set up as that in 2D
(\S\ref{2dvisc}), but with 
\begin{align}
  \qvert = \frac{Q_\mathrm{3D,crit}}{\sqrt{\alpha}}, \label{Q3d_visc_model}
\end{align}
where $Q_\mathrm{3D,crit}\simeq0.36$ is the 3D equivalent to the 2D
critical value, $Q_\mathrm{crit}$. Note that the background vertical
structure now varies with $\alpha$ through Eq. \ref{Q3d_visc_model},
which in turn depends on the cooling time through thermal
equilibrium (Eq. \ref{alpha_beta_relation}). 

\begin{figure}
  \includegraphics[width=\linewidth,clip=true,trim=0cm 0.cm 0.cm
    0.0cm]{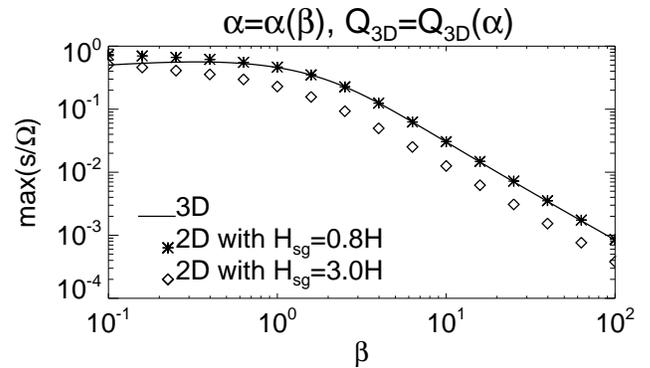}
  \caption{Growth rates from the viscous 3D eigenvalue problem (black solid
    line). Asterisks and diamonds are obtained from the 2D dispersion
    relation (Eq. \ref{thindisk}) with softened gravity as described
    in Appendix \ref{3dcorr}. \label{3d_visc}} 
\end{figure}

Fig. \ref{3d_visc} shows growth rates, maximized over $k$, as a
function of the cooling time $\beta$. We also plot 2D results with 3D
corrections. Softening the self-gravity in 2D captures the correct 
qualitative behavior of the full 3D case. For $\beta\gtrsim 1$
choosing $H_\mathrm{sg}=0.8H$ produces a good match. However, it is
clear that a single, constant value of $H_\mathrm{sg}$ cannot
re-produce 3D growth rates for all $\beta$. This suggests that the
exact value of $H_\mathrm{sg}$ is problem-dependent, although taking 
$H_\mathrm{sg}\sim O(H)$ should give the correct 3D growth rate within 
a factor of two. 

\begin{figure}
  \includegraphics[width=\linewidth,clip=true,trim=0cm 0.cm 0.cm
    0.0cm]{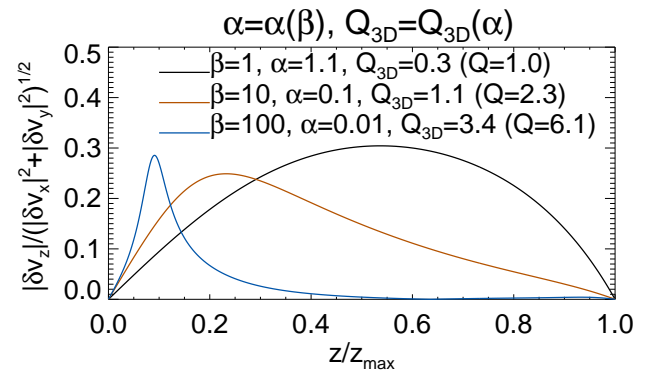}
  \caption{Magnitude of vertical velocities, normalized by the
    magnitude of the total horizontal velocity, of the viscous GI in 
    Fig. \ref{3d_visc}, for three cooling times: $\beta=1,\,10$, and $100$. \label{3d_visc_vz}} 
\end{figure}

Fig. \ref{3d_visc_vz} shows the magnitude of the vertical velocity 
$|\delta v_z|$ scaled by the total horizontal velocity for $\beta =
1,\,10$ and $100$. Vertical speeds are sub-dominant at $\lesssim 30\%$
of the total horizontal speeds. 
These vertical velocities are associated with viscous
  GI, and should not be compared with those associated with the
  underlying gravito-turbulence \citep[e.g.][]{shi14}. Vertical
  velocities are formally neglected in our framework when defining the basic
  state (\S\ref{steady_state}).

%% file: application.tex
\section{Application to protoplanetary disks}\label{2dppd}
We now apply our linear framework to assess the stability of  
PPDs. We consider the gravito-turbulent disk models recently
developed by \citet[][hereafter \citetalias{rafikov15}]{rafikov15}. 
This 2D, Keplerian disk orbits a Solar 
mass star and is defined by the following parameters, 

\begin{itemize}
  \item $\dot{M}$, the global radial mass accretion rate;
  \item $Q_0$, the value of the 2D Toomre parameter where the disk is
    gravito-turbulent;
  \item $T_\mathrm{irr}$, the irradiation temperature;
  \item $\alpha_m$, the dimensionless viscosity associated with other
    sources of turbulence, such as magneto-rotational instabilities
    (MRI, see also \S\ref{visc_gi} and \S\ref{MHD}). 
\end{itemize} 
These properties serve as inputs for calculating thermal equilibrium, and mass and
angular momentum conservation. Together with an opacity law, 
they allow us to construct a global disk model with surface density $\Sigma(R)$ and
temperature $T(R)$ where $R$ is the global cylindrical radius from the
star. \citepalias[See][for details.]{rafikov15} These profiles give $Q(R)$,
required for input into our linear framework. We derive other
dimensionless parameters below. 

Although our disk is global in extent, here we consider the local
stability at each radius. We are thus neglecting any global
instabilities \citep{adams89,lodato05,kratter10} as well as any
evolution of the disk properties in response to GI. 
 

\subsection{Effective $\alpha$}
In a steady, viscously accreting Keplerian disk 
with constant $\dot{M}$ we have,
approximately, $\nu\Sigma = \dot{M}/3\pi$. Hence
\begin{align}
  \alpha = \frac{\dot{M}}{3\pi}\frac{\Omega(R)}{c_{s}^2(T)\Sigma(R)},  
\end{align} 
which sets the viscosity coefficient to be used at each radius. Note that the disk profiles 
employed here give constant $\dot{M}$ rather than constant $\alpha$.

\subsection{PPD beta cooling}\label{ppd_cooling}
Energy loss in \citetalias{rafikov15} is given by 
\begin{align}\label{real_cool}
  \Lambda = \frac{2\sigma}{f(\tau)}\left(T^4 - T_\mathrm{irr}^4\right)
\end{align}
per unit area, where
\begin{align}
  f(\tau) = \tau + \frac{1}{\tau}, \label{ftau} 
\end{align}
and
\begin{align}
  \tau = \kappa_d(T)\Sigma
\end{align}
is the optical depth. Recall Eq. \ref{opacity_law} 
is our opacity model where $\kappa_d\propto T^b$ with $b=2$. 
Eq. \ref{ftau} accounts for cooling in the optically-thin ($\tau\ll
1$) and optically-thick ($\tau\gg1$) regimes. 

Note that Eq. \ref{real_cool} falls within our definition of a beta
cooling prescription, because 
it is an explicit function of the thermodynamic states. However, we
formulated the linear problem with the standard beta cooling function 
given by Eq. \ref{beta_cool}, which has a different (less realistic)
dependence on disk temperature. 
 In order to adapt the existing  framework to the above PPD cooling function, we 
need to identify the equivalent $\beta$ and $\theta$ parameters that are required for the 
linearized equations (see \S\ref{linear_bcool}). 
 

Linearizing Eq. \ref{real_cool} gives   
\begin{align}\label{linear_real_cool}
  (\gamma-1)\frac{\delta\Lambda}{\Sigma} = \frac{2\sigma(\gamma-1)
    T^4C_1}{f(\tau)c_{s}^2\Sigma}\left(\frac{\delta P}{\Sigma} -
  \frac{C_2}{C_1}c_{s}^2\frac{\delta\Sigma}{\Sigma}\right), 
\end{align}
where
\begin{align}
C_1(\tau, T) &= 4 - b\times g(\tau, T),\\ 
C_2(\tau, T) &= 4 + (1-b)\times g(\tau, T),\\ 
  g(\tau, T) &= \left( \frac{\tau^2-1}{\tau^2+1}\right)\left(1 -
  \frac{T_\mathrm{irr}^4}{T^4}\right). \label{g_def}
\end{align}
Comparing Eq. \ref{linear_real_cool} with the linearized form of the
standard cooling function, Eq. \ref{linear_beta}, we identify
\begin{align}
  &\beta = \frac{f(\tau)c_s^2\Sigma\Omega}{2\sigma(\gamma-1)C_1T^4},\label{real_beta}\\
  &\theta = \frac{C_2}{C_1},\label{real_theta} 
\end{align}
to be used in the 2D dispersion relation (Eq. \ref{thindisk}). 
Eq. \ref{real_beta} represents a physical cooling time \emph{for the
  perturbations}, and is consistent with previous definitions within
factors of order unity \citep[e.g.][their Eq. 2]{kratter10}.  

Eq. \ref{real_theta} shows that $\theta$ 
is related to the true irradiation temperature $\tirr$ through the function $g$ given by 
Eq. \ref{g_def}, and is therefore a only a weak function of the irradiation temperature. 
More specifically $\theta = O(1)$ 
for all $\tirr <T$, and for our adopted opacity law, Eq. \ref{opacity_law}, 
\begin{align*}
  \theta = \frac{4-g}{4-2g}. 
\end{align*}
Thus for $T_\mathrm{irr} = 0$ we have $5/6<\theta<3/2$ by considering
$\tau\to 0,\,\infty$. However, for $T=\tirr$ we have $\theta = 1$, as
expected intuitively.    


\subsection{Inviscid stability condition}
With our new linearized cooling function in hand, from the discussion
in \S\ref{2d_inviscid} and by applying 
Eq. \ref{stable_condition},  we conclude that without viscous effects 
the disk is stable everywhere if  
\begin{align} 
  \gamma > \frac{3}{2} \quad \text{and} \quad Q >
  \sqrt{\frac{6}{5}} \label{ppd_invisc_cond} 
\end{align} 
are both satisfied.  

PPDs become irradiation-dominated at large distances from the star,
where $T\to\tirr$ and $\theta\to 1$ \citep{chiang97,dalessio97,kratter11}.
Then Eq. \ref{ppd_invisc_cond} relaxes to 
$\gamma,\, Q > 1$ in the outer disk. The condition on $\gamma$
is then guaranteed. On the other hand, numerical 
simulations of gravito-turbulence show that $1\lesssim Q \lesssim 2$
\citep{gammie01,rice11}, and the second inequality is generally
satisfied.   
Taken together, this suggests that in the outer regions of a realistic
PPD, cooling 
may not be the primary cause for a secondary 
instability of a gravito-turbulent disk, leading to
fragmentation. This leaves viscous GI as the only possible culprit
within our framework, as we illustrate below.  

\subsection{Example 2D calculation}\label{pp2d_example}
\begin{figure}
  \includegraphics[width=\linewidth,clip=true,trim=0cm 0cm 0cm
    0.0cm]{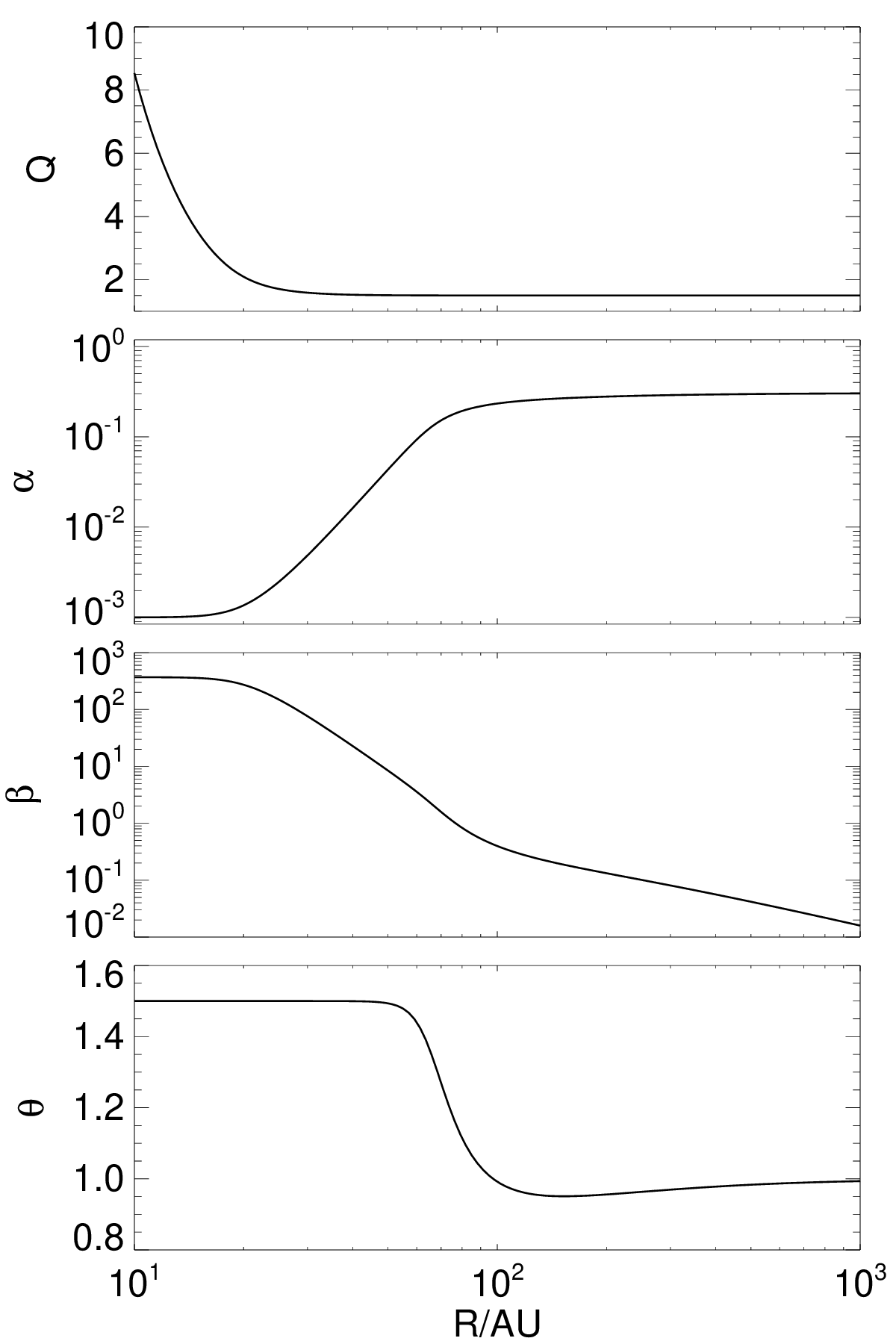}
  \caption{Equilibrium profile obtained from the disk model developed
    by \cite{rafikov15}, with parameters $\dot{M} =
    10^{-6}M_\sun\yr^{-1}$, $Q_0=1.5$, $\tirr=10\mathrm{K}$, and
    $\alpha_m=10^{-3}$.   
    \label{rafikov_model}}
\end{figure}

We relax the inviscid assumption and consider a fiducial disk  model with
$\dot{M} = 10^{-6}M_\sun\,\yr^{-1}$, $Q_0=1.5$,  
$\tirr=10\mathrm{K}$, and $\alpha_m=10^{-3}$. 
Such a high accretion rate is consistent with those expected for 
young protostellar disks \citep{shu77,enoch09}. Similarly,
$10\mathrm{K}$ is a conservatively low background irradiation level
consistent with cloud temperatures in star forming regions
\citep{plume97,johnstone01}. 
Stellar irradiation will typically elevate $\tirr$ in addition to
adding a radial dependence \citep{kratter08} 
We adopt the opacity scale $\kappa_{d0} =
5\times10^{-4}\mathrm{cm}^2\,\mathrm{g}^{-1}\,\mathrm{K}^{-2}$  as in
\citetalias{rafikov15}. We use $\gamma=1.6$, approximately applicable
to an ideal molecular gas at low temperatures 
This choice of $\gamma$ satisfies the global inviscid stability condition
(Eq. \ref{ppd_invisc_cond}). 
Fig. \ref{rafikov_model} shows the equilibrium 
disk profile in terms of $Q$, $\alpha$, $\beta$, and $\theta$.  
These profiles serve as input to the 2D dispersion relation
(Eq. \ref{thindisk}, Eq. \ref{bigA}---\ref{bigF}). 

\begin{figure}
  \includegraphics[width=\linewidth,clip=true,trim=0cm 2cm 0cm
    0.0cm]{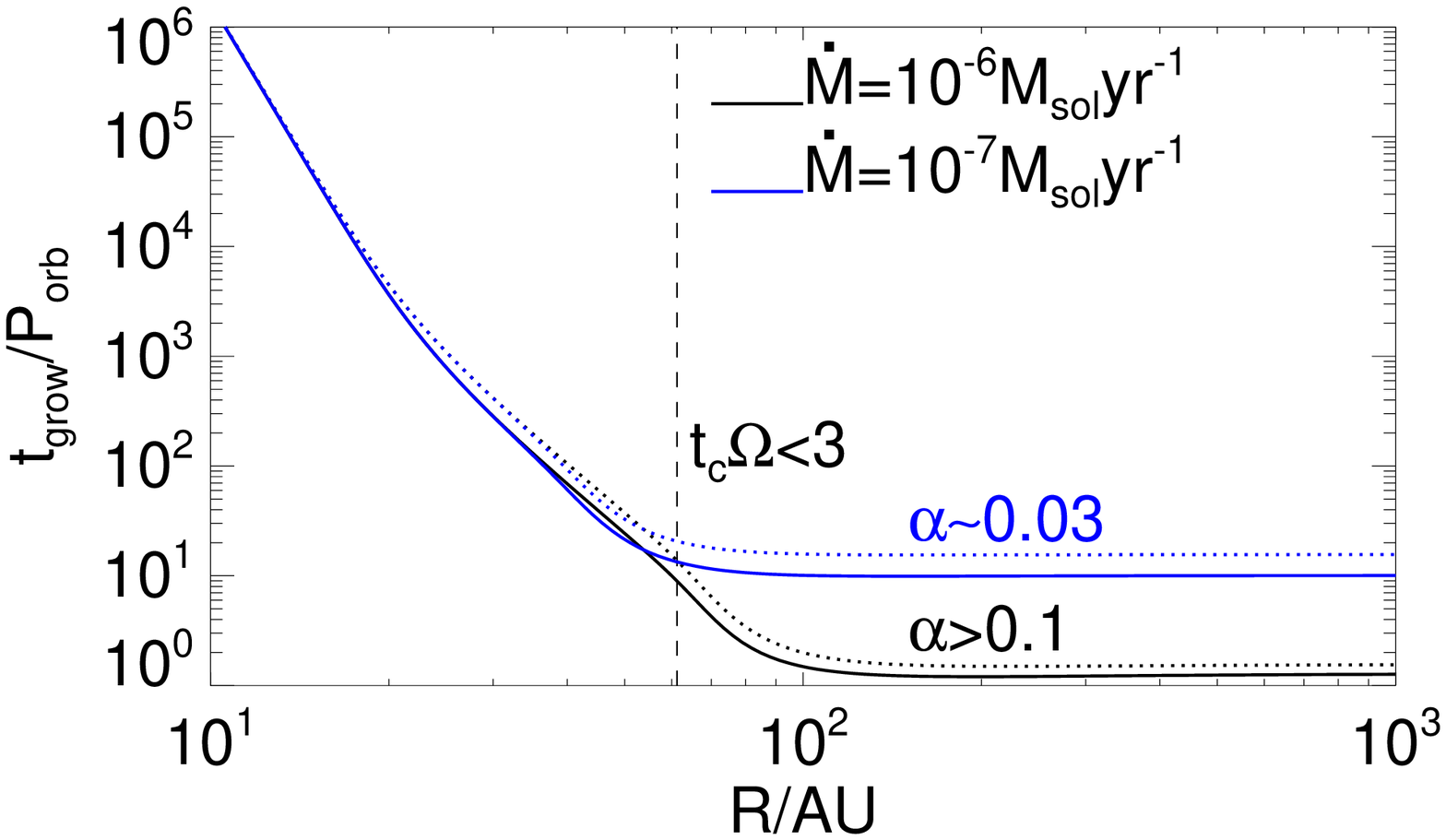}\\
  \includegraphics[width=\linewidth,clip=true,trim=0cm 0cm 0cm
    0.cm]{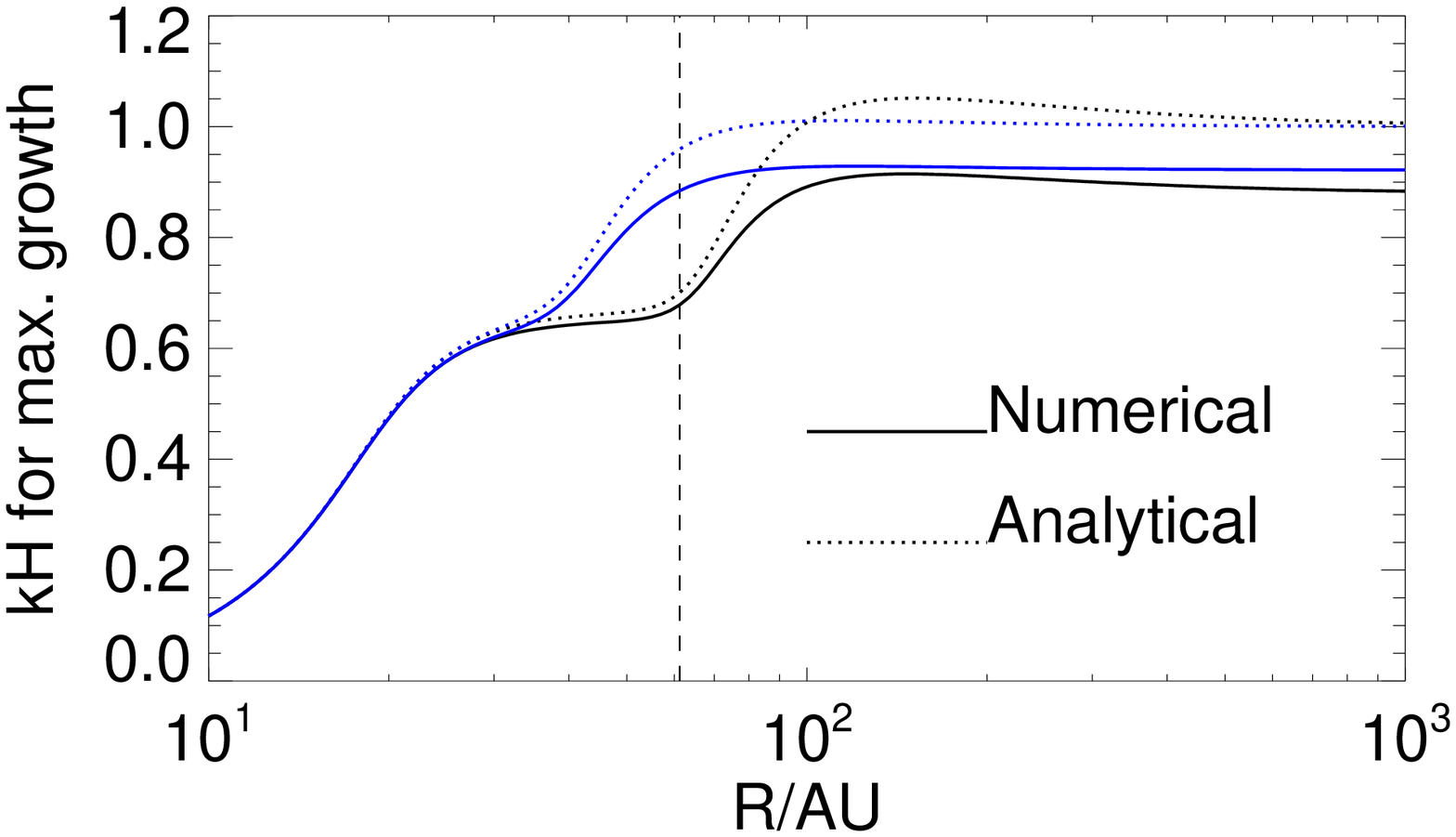}
  \caption{Black lines show growth timescales (top) of the most
    unstable wavenumber (bottom) for viscous, 
    self-gravitational modes in the 2D PPD model shown in
    Fig. \ref{rafikov_model}. Blue curves are for the same disk
    model but with a lower accretion rate. Solid curves are obtained
    numerically from Eq. \ref{thindisk}, and dotted curves 
    are analytic results based on Eq. \ref{gammie_smallk}. 
      For both accretion rates the cooling time $\lesssim 3/\Omega$
      beyond the vertical dashed line, but for the 
      $\dot{M}=10^{-6}M_\sun\mathrm{yr}^{-1}$ disk $\alpha > 0.1$
      beyond this radius, while $\alpha\lesssim
      0.03$ throughout the
      $\dot{M}=10^{-7}M_\sun\mathrm{yr}^{-1}$ disk. 
    \label{rafikov_growth}}
\end{figure}

Fig. \ref{rafikov_growth} shows growth timescales and
optimum wavenumbers for viscous GI in this fiducial model. For
comparison we also plot a case with lower accretion rate, 
$\dot{M}=10^{-7}M_\sun\,\yr^{-1}$; and analytic estimates based on
Eq. \ref{gammie_smallk} (instead of Eq. \ref{gammie_maxrate_simple}
since here $\theta\sim 1$) which gives the optimum wavenumber and
growth rates as 
\begin{align}
  |K| = \frac{3}{2\theta Q}, \quad
  S = \frac{27\alpha}{16\theta^3Q^4}. 
\end{align}
These are similar to the isothermal results of
\citet[][their Eq. 19 and 21, respectively]{sterzik95}, and identical if one
takes $\theta=1$. 
 
The most unstable wavelength is a few times the disk thickness. 
So long as $H\ll R$, this result is consistent with our use of the
local approximation. For our fiducial disk, $H/R \sim 0.07$ around 
$R\sim 100$AU, and $H/R <0.25$ throughout the disk. 
The increase in $|K|$ from 
$\sim 10\mathrm{AU}$ to $\sim 20\mathrm{AU}$ is due to the decrease in
$Q$, while that from $\sim 
60\mathrm{AU}$ to $\sim 100\mathrm{AU}$ occurs as the disk transitions
from the 
optically-thick to optically-thin regime. The mismatch between the
numerical and analytic solutions at large
distances is expected since the above expressions assume $|K|\ll
1$. Nevertheless the analytic estimates reproduce qualitatively correct
behavior. 

The fiducial disk is subject to viscous GI on dynamical timescales
($\lesssim 10$ orbits) for $R\gtrsim60$AU. We note this transition
radius is also implied by Eq. 16 in \cite{kratter10}. Coincidentally,
beyond this radius $\alpha\gtrsim 0.1$ (and 
$\tcool\Omega\lesssim 3$), which is often quoted as a condition for disk
fragmentation   
\citepalias[e.g.][]{rafikov15}. Thus viscous GI may be responsible for 
the transition between gravito-turbulence and fragmentation due to the
removal of rotational support by viscous (turbulent) stresses. 

On the other hand, the lower $\dot{M}$ model also attain $\tcool\lesssim
3\Omega^{-1}$ beyond $\sim 60$AU, but $\alpha\lesssim 0.03$
everywhere. 
Applying empirical cooling conditions for fragmentation may then lead to
contradiction. Instead, if viscosity is the physical cause for
fragmentation, then our result suggest the lower $\dot{M}$ disk should
not fragment (at least much less likely than our fiducial case) 
because the instability cannot develop on orbital timescales. 

\subsection{3D PPD with radiative diffusion}
We briefly consider 3D PPDs with explicit radiative diffusion
(\S\ref{rad_cool}). 
Given the $\alpha(R)$ and $Q(R)$ profiles obtained
from the 2D model above, at each radius $R$ we obtain the vertical
structure from Eq. \ref{vert_eq1}---\ref{thermal_eq}, with 
Eq. \ref{rad_cool1}---\ref{rad_cool2} for the radiative flux.
We then solve the 3D eigenvalue problem as in \S\ref{3ddisk}, with the
additional boundary condition that the disk surface temperature is
fixed, $\delta T(\zmax) = 0$. 
We use the fiducial disk model as in \S\ref{pp2d_example} but 
with $T_\mathrm{irr}=0$, since our simple radiative diffusion
treatment does not include irradiation (\S\ref{rad_cool}). 
We use a slightly smaller vertical
domain with $\rho(\zmax)=0.1\rho_0$.  

\begin{figure}
  \includegraphics[width=\linewidth,clip=true,trim=0cm 2cm 0cm
    0.0cm]{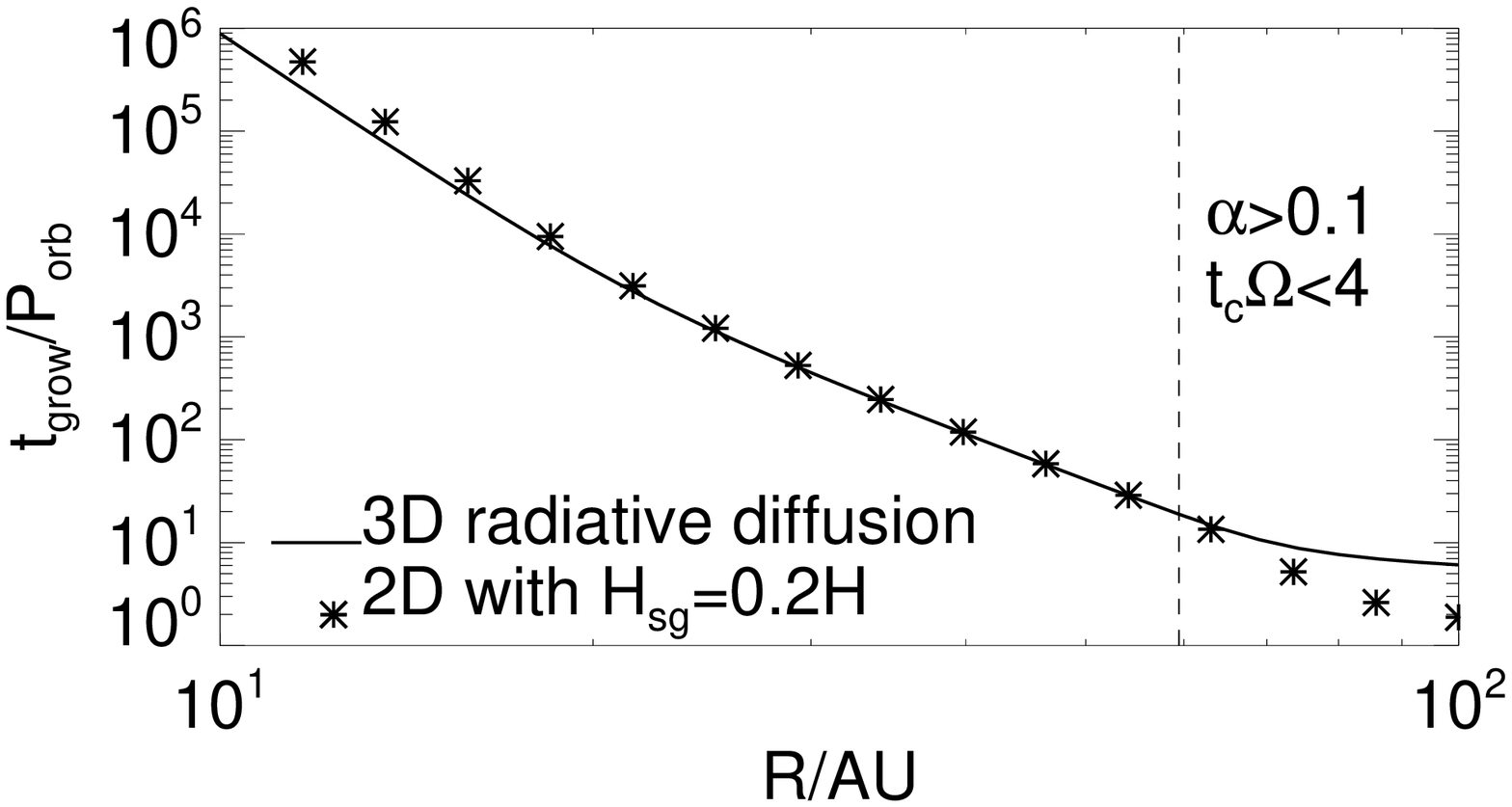}\\
  \includegraphics[width=\linewidth,clip=true,trim=0cm 0cm 0cm
    0.8cm]{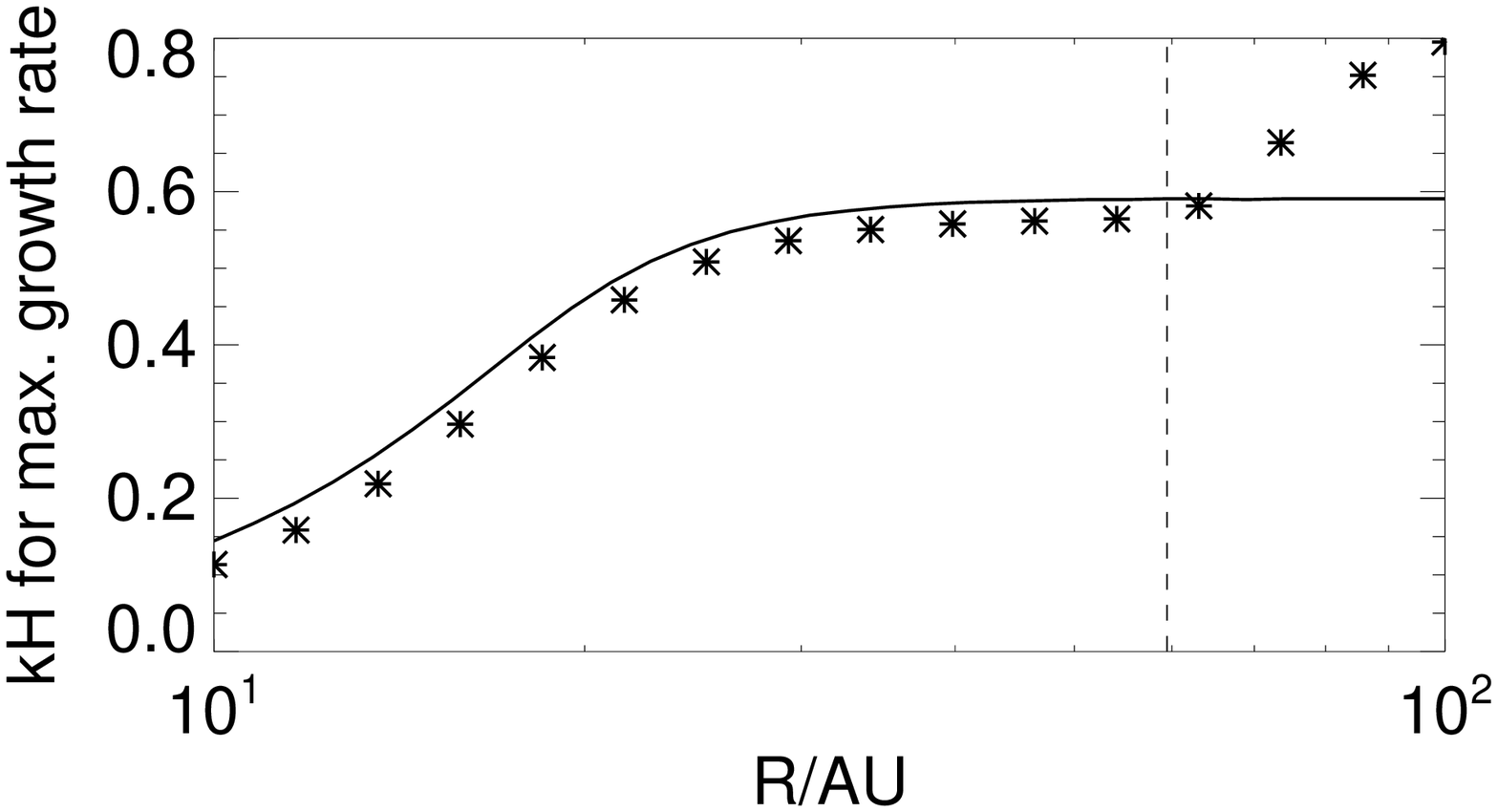}
  \caption{Growth timescales (top) of the most unstable
    wavenumber (bottom) for viscous GI in a 3D PPD with radiative
    diffusion  (black lines). Asterisks are corresponding results
    obtained from the corresponding 2D problem with softened gravity.   
      Beyond the vertical dashed line $\alpha>0.1$, which corresponds to
      cooling times $t_\mathrm{c}\lesssim 4/\Omega$.  
    \label{rafikov_growth3d}}
\end{figure}

Fig. \ref{rafikov_growth3d} shows the growth rates and most unstable 
wavenumber for $R\in[10,100]$AU; along with the corresponding 2D 
results matched with softened self-gravity. There is good agreement for
$R\lesssim60$AU where the disk is optically thick ($\tau\gtrsim
1$) and thus both models apply. However, beyond $60$AU where the disk
becomes optically-thin, radiative diffusion (the 3D curve) is not
valid and under-estimates the growth rates. 
Nevertheless, the transition radius of  
$\sim60$AU, beyond which growth timescales become dynamical,    
can be correctly calculated within the 2D framework.  

%% file: summary.tex
\section{Summary and discussion}\label{summary}
In this paper, we develop the linear theory of cooling, irradiated, 
and viscous accretion disks in order to understand gravitational 
instability (GI) in realistic protoplanetary disks (PPDs). 
We use a Navier-Stokes viscosity to mimic the effects of
turbulent angular momentum transport. 
This viscosity provides a background heating to balance the imposed cooling,  
and may also act on linear perturbations. 
We suggest that disk fragmentation observed in numerical simulations  
can be understood as the eventual outcome of secondary
instabilities of a gravito-turbulent base state, driven by cooling
and/or viscosity. 

Previous work has focused on the impact of viscosity and 
cooling on the equilibrium temperature and surface density of an 
accretion disk, but merely used this to calculate  the classic Toomre $Q$,
thereby assess stability. 
We demonstrate by explicitly including these effects into the
dispersion relation that they can drive secondary instabilities.
While viscosity and cooling can be related through thermal
balance, they independently enhance growth rates: cooling reduces
thermal stabilization; and viscous forces compromise rotational
stabilization. This provides a physical explanation as to why
rapidly-cooled, gravito-turbulent disks cannot exist. Moreover, we discuss below how
these models may lend support to the varied behavior observed in numerical simulations.


The effect of cooling and irradiation on GI is quantified by the 
dispersion relation Eq. \ref{inviscid}. 
We find sufficient conditions for stability which  
depends on the irradiation level (Eq. \ref{stable_condition}) but is 
\emph{independent} of the cooling time. 
This means that long cooling times can still  
formally lead to instability. However, growth timescales may be 
uninterestingly long for cooling times $\tcool\Omega\gtrsim 
O(10)$. Because cooling affects pressure support, GI driven by cooling
occur on small scales, $kH\gtrsim O(1)$.   


We generalize the `viscous gravitational instability', previously 
studied in isothermal disks
\citep{lynden-bell74,willerding92,gammie96}, to include 
cooling, viscous heating and irradiation. 
We consider a disk with viscosity and self-gravity inversely
related ($\alpha\propto Q^{-2}$) to model a 
gravito-turbulent background, and find viscous GI occurs on orbital
timescales for $\alpha\gtrsim 0.1$. This is consistent with the notion 
of a maximum stress sustainable by gravito-turbulence established 
by numerical simulations \citep{rice05}. Because viscosity affects 
rotational support, viscous GI occurs on large scales, $kH\lesssim
O(1)$.  Furthermore, irradiation preferentially
stabilizes small-scale perturbations. 
                             

We apply our linear framework to protoplanetary disks with 
realistic models for cooling and gravito-turbulence. 
We show that with a physically motivated cooling model for PPDs, 
cooling alone does not lead to gravitational instabilities. This is
due to stabilization by an effective `irradiation' associated with the 
density-dependence of the PPD cooling function as it appears in the
stability problem, which is present even if there is no physical
irradiation. This captures the fact that density enhancements impede cooling. 
 




Instead, viscous GI occur on dynamical timescales in a PPD for
$R\gtrsim 60$AU because $\alpha\gtrsim0.1$ there. 
This corresponds to a Toomre $Q\simeq 1.5$ and 
a cooling time $\tcool\lesssim 3\Omega^{-1}$. These are coincident with 
\emph{empirical} conditions cited in the literature to determine disk
fragmentation 
\citep[e.g.][]{rafikov15}. Here, we attribute a \emph{physical}
cause for the fragmentation of realistic PPDs: gravito-turbulent PPDs
fragment when turbulent stresses are large enough to further
destabilize the disk against self-gravity.    


\subsection{Relation to numerical simulations}\label{prev_works}
Our results may help understand some numerical simulations 
concerning disk fragmentation. Table 
\ref{bstar_compare} shows a close match between the characteristic
cooling time for cooling-driven GI (Eq. \ref{betastar}) 
and that for disk fragmentation observed in simulations 
\citep{gammie01,rice05,rice11}. This suggests that, at least for those
simulations, fragmentation is physically due to the removal of
thermal stabilization by cooling on radial lengthscales of the disk 
thickness. In this interpretation, gravito-turbulence only
provides a background heating. 
Our characteristic cooling time corresponds to a dimensionless background viscosity 
as defined in the above studies\footnote{This differs from our
  definition of $\alpha$ by a factor of $\gamma^{-1}$.}  
\begin{align}
  \alpha =
  \frac{4}{9}\frac{\left(\sqrt{\gamma}-1\right)^{1/2}}{\gamma\left(\sqrt{\gamma}+1\right)}
  \simeq \begin{cases}
    0.062 & \gamma = 7/5, \\
    0.063 & \gamma = 5/3,\\
    0.059 & \gamma = 2.
  \end{cases}
\end{align}
This $\alpha\sim0.06$ is roughly constant, consistent with \cite{rice05}. 





More recent simulations have raised the issue of numerical
convergence. 
\cite{meru11} found that better resolved disks fragmented at longer cooling times.
Follow-up studies attributed at least some of this effect to decreasing numerical viscous
heating at higher resolution, which helps fragmentation
\citep{lodato11,meru12}. 
We can expect this if numerical viscosity
contributes to an effective irradiation, 
because then perturbations can cool to lower temperatures with
increasing resolution, see Fig. \ref{invisc_theta}.
  In global simulations, non-convergence has also been attributed to initial 
  conditions that lead to internal edges \citep{paardekooper11b}, but
  this cannot be modeled in our local setup. 
  \cite{rice12} point out that the standard implementation of beta
  cooling in smoothed-particle hydrodynamics (SPH) applies cooling on scales well-below the SPH smoothing
  lengths, and that this inconsistency may contribute to
  non-convergence. 

 However, \cite{paardekooper12} also found in local 2D
  grid-based simulations that fragmentation can occur  
for  slowly-cooled disks with $\tcool\Omega\gg O(1)$, but that this requires   
simulations to run for significantly longer than dynamical 
timescales. This is consistent with our finding
that for either cooling-driven or viscous GI, there is no critical 
cooling rate/viscosity below  which the disk is  
formally stable. Instead, growth rates smoothly decrease with
increasing $\tcool$ (decreasing $\alpha$), implying that
instabilities, and hence fragmentation, simply take longer to develop
for slowly-cooled disks. 
  However, to properly consider long timescales, it may be necessary
  to account for
  secular evolution in the global disk. 



Here, we highlight that most numerical experiments, including those
above, employ the standard beta cooling function, Eq. \ref{beta_cool},
\emph{without} a physical floor temperature. We show in \S\ref{2d_inviscid}
(see also \S\ref{cool_gi}) that this implies cooling-driven GI can
occur at any sufficiently small scale. Therefore as the numerical
resolution increases, simulations can access a wider range of unstable
scales. Although small-scale modes have weaker growth rates, they can 
become important over long timescales.   
In this respect, it is perhaps not surprising to find non-convergence with 
increasing resolution and/or integration times. 
  On the other hand, the convergence issue may be
  less serious in 3D since small-scale modes are more stable in 3D
  than in 2D (Appendix \ref{3dcorr}). 

We suggest having a physical floor temperature in the standard beta
cooling prescription is  necessary for  
numerical convergence. This limits the
relevant scales to a finite range.  
Furthermore, without a floor temperature, standard beta cooling is
a function of the pressure/energy density field only. 
There may be some inconsistency in applying results obtained from 
this to actual PPDs where cooling depends on two thermodynamic states (e.g. pressure and
temperature). A floor temperature permits a mapping between standard beta
cooling and PPD cooling (\S\ref{ppd_cooling}). 
Moreover, the standard beta cooling does 
not account for optical depth effects, which force the mid-plane 
and high density perturbations to cool more slowly.

%


A temperature floor may be a necessary, but not sufficient condition
for numerical convergence. \cite{baehr15} included a floor temperature 
in their local 2D simulations but still find that at fixed cooling rates, disks
eventually fragment with sufficient spatial resolution.
In light of our results on viscous GI, 
we suggest another possible contribution to non-convergence: 
high resolution enables small-scale turbulent angular momentum transport 
to aid clump formation via the removal of rotational support. (See also \S\ref{MHD} below.) 
If clumps are only marginally resolved, simulations may not have sufficient dynamic range for turbulent eddies
to cascade down to these scales.

We comment that although modern simulations resolve the dominant
scale associated with gravito-turbulence very well, $kH\sim 1$
\citep{cossins09}, this does not
necessarily imply small-scale dynamics/thermodynamics are unimportant 
\citep[especially for non-linear evolution, ][]{young15},  
as is evident from the non-converging simulations described above.   
Even at very high resolution one might worry that the artificial dissipation scale
imposed by the grid/smoothing length is too similar to the scale of fragmentation.

\subsection{MHD turbulence}\label{MHD}
We emphasize that the linear framework we have developed does not assume  
a particular origin for the turbulence that is represented by the imposed
viscosity. For example, our 2D dispersion 
relation, Eq. \ref{thindisk} with Eq. \ref{bigA}---\ref{bigF}, treats 
$\alpha$ as an independent input parameter. 


Our models may thus apply to self-gravitating disks dominated by MHD
turbulence.     
Explicit numerical simulations of magnetized, massive disks have been performed 
by \cite{fromang05}. This study finds disk fragmentation with increasing numerical
resolution, and attributes this to resolving the most unstable MRI
wavelength, which enables small-scale angular momentum removal by MHD
turbulence to aid fragmentation. 
This physical mechanism is represented by the viscous GI 
discussed in this paper, which lends some support for  
the use of a viscosity to represent turbulence.    


\subsection{Outstanding issues}
True disk fragmentation is a non-linear process characterized by
clumps reaching densities that are orders of magnitude above the
ambient value. They must also survive 
disruption by tidal shear and shocks \citep{shlosman87,young16}.  
Clearly, our linear models cannot address fragmentation
directly. Technically, we have only demonstrated that a 
gravito-turbulent state becomes dynamically unstable, and thus should
not persist, when cooling is too rapid or when the associated viscous
stresses are too large.  However, steady gravito-turbulence or 
fragmentation are the \emph{only} possible outcomes of cooling, 
self-gravitating disks in the local limit \citep[][]{gammie01}.    
Thus it seems reasonable to speculate that the non-existence of a stable
gravito-turbulent state, here due to dynamical instability, implies
disk fragmentation.


Our deterministic approach cannot model `stochastic fragmentation'
\citep{paardekooper12,hopkins13}. In this interpretation,  
fragmentation is attributed to the occurrence and survival 
of large, non-linear density enhancements, which arise from the
gravito-turbulent fluctuations simply by chance. 
There is insufficient evidence that gravito-turbulence adequately samples
the density power spectrum as assumed by \cite{hopkins13}.
Nevertheless, one might consider this form of fragmentation as a secondary instability
triggered by lowering the local 
Toomre parameter through a (random) increase in density. 


The most important assumption in this work is modeling turbulence as a  
Navier-Stokes viscosity. Furthermore, we have chosen a particular  
viscosity law (see \S\ref{visc_model}) to mimic the effect of 
turbulence in reducing rotational support (in the sense that it
provides small-scale angular momentum transport). 
How to quantitatively model the effect of gravito- or MHD turbulence  
as a viscosity, especially on dynamical timescales, should be 
clarified with direct numerical simulations. The present viscosity
models should then be modified accordingly. 

Another possible generalization is non-axisymmetric
disturbances. In barotropic, inviscid disks  
non-axisymmetric global GI can develop for $Q$ somewhat larger than unity
\citep{papaloizou89,adams89,papaloizou91,laughlin97}. It would be interesting to study
how non-axisymmetric perturbations are affected by cooling and
viscosity in order to improve 
the link between disk fragmentation and the
stability of gravito-turbulent disks.   



%% file: appendix3d.tex
\section{2D dispersion relation}\label{2ddisp}
The functions in Eq. \ref{thindisk} are 
\begin{align}
  A(S,K) =& \left(\frac{4}{3}\alpha+\alpha_b\right) K^2 + S +
  \mathcal{E}\mathcal{F}\label{bigA}
  - \frac{2|K|}{QS}, \\
  B(S,K) =& 2\left(\alpha q \mathcal{F} - 1\right),\\
  C(S,K) =& (2 - q) + \frac{\alpha q K^2(1+\mu)}{S} 
  + \alpha q \lambda \mathcal{E}\mathcal{F},\\
  D(S,K) = & \alpha K^2 + S + 2\alpha^2q^2\lambda\mathcal{F},
\end{align}
with
\begin{align}
  \mathcal{E} = \frac{\alpha q^2(1+\mu)}{S} +
    \frac{\gamma}{\gamma-1} + \frac{\theta}{\beta S(\gamma-1)},\quad
  \mathcal{F} = \frac{K^2(\gamma-1)\beta}{1 + \beta S - \alpha\beta
    q^2\lambda(\gamma-1)}\label{bigF}. 
\end{align}
Note that these equations treat all the parameters as independent
(namely $\alpha$, $\alpha_b$, $\beta$, $\theta$, and $Q$).  
 

\section{2D viscous GI} \label{gammie_check}
We obtain the dispersion relation for viscous GI described by
previous authors \citep{lynden-bell74,willerding92,gammie96} as 
follows. We set $\mu=-1, \lambda=0$ in Eq. \ref{visc_law} to obtain the same viscosity
models. Next we consider $|\beta S|\to \infty$, i.e. no explicit
cooling on the perturbations. Then the condition $AD = BC$ implies 
\begin{align}
  S^3 + \left(\frac{7}{3}\alpha + \alpha_b\right)K^2S^2 + \left[2(2-q) -
    \frac{2|K|}{Q} + \gamma K^2 + \alpha K^4 \left(\frac{4}{3}\alpha +
    \alpha_b\right)\right]S + \alpha K^2 \left[\gamma K^2 -
    \frac{2|K|}{Q} - 2q(2-q)(\gamma-1)\right]=0,
\end{align}
which agrees with the above studies in the isothermal
limit \citep[$\gamma=1$; see also][their Eq. 28]{schmit95}.  
The non-isothermal term $\propto (\gamma-1)$
originates from viscous dissipation, which was excluded in the
aforementioned works. Its effect is to increase the maximum wavenumber 
for viscous GI. 

An approximate solution for the growth rate may be obtained for small
$|K|$ by balancing the last two terms,
\begin{align}
  S \simeq \frac{\alpha K^2\left[2|K|/Q- \gamma K^2 + 
     2q(2-q)(\gamma-1)\right]}{2(2-q)  - 2|K|/Q + \gamma K^2 }, 
\end{align}
which coincides with \citeauthor{gammie96}'s Eq. 18 for $\gamma=1$. 




\section{3D corrections in 2D theory}\label{3dcorr}
The simplest way to mimic the effect of finite disk thickness
on gravitational instabilities is to weaken self-gravity by reducing 
the gravitational constant  
\begin{align}
  G \to G\left(1+|k|H_\mathrm{sg}\right)^{-1}, 
\end{align}
or equivalently $Q\to Q\left(1+|k|H_\mathrm{sg}\right)$. This
prescription, derived by \cite{shu84}, is widely applied
\citep[e.g.][]{youdin11,takahashi14}. Here $H_\mathrm{sg}$ is a
measure of the disk thickness. We intuitively expect
$H_\mathrm{sg}\sim H$, but its precise value is not known a priori.   
We regard $H_\mathrm{sg}$ as a free parameter of the problem.

\section{Numerical method for the 3D eigenvalue problem}\label{3d_method}
We use a pseudo-spectral method to solve the set of ordinary
differential equations, Eq. \ref{lin_mass}---\ref{lin_gravity}, on 
the domain $ z\in[0,\zmax]$ with a parity condition at the mid-plane.
We expand $\bm{U}=[\delta P,\delta\rho,\delta\Phi,
  \delta v_x, \delta 
  v_y]$ in even Chebyshev polynomials,
\begin{align}
  \bm{U}(z) = \sum_{j=1}^N \bm{a}_jT_{2(j-1)}(z/\zmax), 
\end{align}
and the vertical velocity in odd Chebyshev polynomials,
\begin{align}
  \delta v_z(z) = \sum_{j=1}^N b_jT_{2j-1}(z/\zmax),   
\end{align}
where $\bm{a}_j$ and $b_j$ are the spectral coefficients. 
The basis functions are chosen to satisfy a reflecting boundary
condition at the mid-plane, 
\begin{align}
  \bm{U}^\prime(0) = \bm{0}, \quad \delta v_z(0) = 0.
\end{align}
For simplicity we apply a reflecting upper disk boundary, 
\begin{align}
  \delta v_z(\zmax) = \delta v_x^\prime(\zmax) = \delta
  v_y^\prime(\zmax) = 0,
\end{align}
and the potential satisfies
\begin{align}
  \delta \Phi^\prime (\zmax) + k \delta\Phi(\zmax) = 0, 
\end{align}
as derived by \cite{goldreich65a} and used in similar studies  
\citep{kim12,lin14c}. 

We discretize the equations, including upper disk boundary conditions,
over the $N$ positive abscissae of the extrema of $T_{2N-1}$. This
procedure converts the differential equations into a generalized
eigenvalue problem, for which we use the standard matrix package
LAPACK to solve. We use $N=65$.

%% file: paper3d.bbl
\begin{thebibliography}{}
\expandafter\ifx\csname natexlab\endcsname\relax\def\natexlab#1{#1}\fi

\bibitem[{{Adams} {et~al.}(1989){Adams}, {Ruden}, \& {Shu}}]{adams89}
{Adams}, F.~C., {Ruden}, S.~P., \& {Shu}, F.~H. 1989, \apj, 347, 959

\bibitem[{{Armitage}(2011)}]{armitage10}
{Armitage}, P.~J. 2011, \araa, 49, 195

\bibitem[{{Armitage} {et~al.}(2001){Armitage}, {Livio}, \&
  {Pringle}}]{armitage01}
{Armitage}, P.~J., {Livio}, M., \& {Pringle}, J.~E. 2001, \mnras, 324, 705

\bibitem[{{Baehr} \& {Klahr}(2015)}]{baehr15}
{Baehr}, H., \& {Klahr}, H. 2015, \apj, 814, 155

\bibitem[{{Balbus} \& {Papaloizou}(1999)}]{balbus99}
{Balbus}, S.~A., \& {Papaloizou}, J.~C.~B. 1999, \apj, 521, 650

\bibitem[{{Bell} \& {Lin}(1994)}]{bell94}
{Bell}, K.~R., \& {Lin}, D.~N.~C. 1994, \apj, 427, 987

\bibitem[{{Boss}(1997)}]{boss97}
{Boss}, A.~P. 1997, Science, 276, 1836

\bibitem[{{Chiang} \& {Goldreich}(1997)}]{chiang97}
{Chiang}, E.~I., \& {Goldreich}, P. 1997, \apj, 490, 368

\bibitem[{{Clarke} {et~al.}(2007){Clarke}, {Harper-Clark}, \&
  {Lodato}}]{clarke07}
{Clarke}, C.~J., {Harper-Clark}, E., \& {Lodato}, G. 2007, \mnras, 381, 1543

\bibitem[{{Cossins} {et~al.}(2009){Cossins}, {Lodato}, \& {Clarke}}]{cossins09}
{Cossins}, P., {Lodato}, G., \& {Clarke}, C.~J. 2009, \mnras, 393, 1157

\bibitem[{{D'Alessio} {et~al.}(1997){D'Alessio}, {Calvet}, \&
  {Hartmann}}]{dalessio97}
{D'Alessio}, P., {Calvet}, N., \& {Hartmann}, L. 1997, \apj, 474, 397

\bibitem[{{Enoch} {et~al.}(2009){Enoch}, {Evans}, {Sargent}, \&
  {Glenn}}]{enoch09}
{Enoch}, M.~L., {Evans}, N.~J., {Sargent}, A.~I., \& {Glenn}, J. 2009, \apj,
  692, 973

\bibitem[{{Fromang}(2005)}]{fromang05}
{Fromang}, S. 2005, \aap, 441, 1

\bibitem[{{Gammie}(1996)}]{gammie96}
{Gammie}, C.~F. 1996, \apj, 457, 355

\bibitem[{{Gammie}(2001)}]{gammie01}
---. 2001, \apj, 553, 174

\bibitem[{{Goldreich} \& {Lynden-Bell}(1965)}]{goldreich65}
{Goldreich}, P., \& {Lynden-Bell}, D. 1965, \mnras, 130, 125

\bibitem[{{Goldreich} \& {Lynden-Bell}(1965)}]{goldreich65a}
{Goldreich}, P., \& {Lynden-Bell}, D. 1965, \mnras, 130, 97

\bibitem[{{Goodman} \& {Pindor}(2000)}]{goodman00}
{Goodman}, J., \& {Pindor}, B. 2000, Icarus, 148, 537

\bibitem[{{Helled} {et~al.}(2014){Helled}, {Bodenheimer}, {Podolak}, {Boley},
  {Meru}, {Nayakshin}, {Fortney}, {Mayer}, {Alibert}, \& {Boss}}]{helled14}
{Helled}, R., {Bodenheimer}, P., {Podolak}, M., {et~al.} 2014, Protostars and
  Planets VI, 643

\bibitem[{{Hopkins} \& {Christiansen}(2013)}]{hopkins13}
{Hopkins}, P.~F., \& {Christiansen}, J.~L. 2013, \apj, 776, 48

\bibitem[{{Hunter} \& {Horak}(1983)}]{hunter83}
{Hunter}, Jr., J.~H., \& {Horak}, T. 1983, \apj, 265, 402

\bibitem[{{Johnson} \& {Gammie}(2003)}]{johnson03}
{Johnson}, B.~M., \& {Gammie}, C.~F. 2003, \apj, 597, 131

\bibitem[{{Johnstone} {et~al.}(2001){Johnstone}, {Fich}, {Mitchell}, \&
  {Moriarty-Schieven}}]{johnstone01}
{Johnstone}, D., {Fich}, M., {Mitchell}, G.~F., \& {Moriarty-Schieven}, G.
  2001, \apj, 559, 307

\bibitem[{{Kim} {et~al.}(2012){Kim}, {Kim}, {Seo}, \& {Hong}}]{kim12}
{Kim}, J.-G., {Kim}, W.-T., {Seo}, Y.~M., \& {Hong}, S.~S. 2012, \apj, 761, 131

\bibitem[{{Kimura} \& {Tsuribe}(2012)}]{kimura12}
{Kimura}, S.~S., \& {Tsuribe}, T. 2012, PASJ, 64, 116

\bibitem[{{Kratter} \& {Lodato}(2016)}]{kratter16}
{Kratter}, K.~M., \& {Lodato}, G. 2016, ArXiv e-prints, arXiv:1603.01280

\bibitem[{{Kratter} \& {Matzner}(2006)}]{kratter06}
{Kratter}, K.~M., \& {Matzner}, C.~D. 2006, \mnras, 373, 1563

\bibitem[{{Kratter} {et~al.}(2008){Kratter}, {Matzner}, \&
  {Krumholz}}]{kratter08}
{Kratter}, K.~M., {Matzner}, C.~D., \& {Krumholz}, M.~R. 2008, \apj, 681, 375

\bibitem[{{Kratter} \& {Murray-Clay}(2011)}]{kratter11}
{Kratter}, K.~M., \& {Murray-Clay}, R.~A. 2011, \apj, 740, 1

\bibitem[{{Kratter} {et~al.}(2010){Kratter}, {Murray-Clay}, \&
  {Youdin}}]{kratter10}
{Kratter}, K.~M., {Murray-Clay}, R.~A., \& {Youdin}, A.~N. 2010, \apj, 710,
  1375

\bibitem[{{Latter} \& {Ogilvie}(2006)}]{latter06}
{Latter}, H.~N., \& {Ogilvie}, G.~I. 2006, \mnras, 372, 1829

\bibitem[{{Lau} \& {Bertin}(1978)}]{lau78}
{Lau}, Y.~Y., \& {Bertin}, G. 1978, \apj, 226, 508

\bibitem[{{Laughlin} {et~al.}(1997){Laughlin}, {Korchagin}, \&
  {Adams}}]{laughlin97}
{Laughlin}, G., {Korchagin}, V., \& {Adams}, F.~C. 1997, \apj, 477, 410

\bibitem[{{Laughlin} \& {Rozyczka}(1996)}]{laughlin96b}
{Laughlin}, G., \& {Rozyczka}, M. 1996, \apj, 456, 279

\bibitem[{{Levermore} \& {Pomraning}(1981)}]{levermore81}
{Levermore}, C.~D., \& {Pomraning}, G.~C. 1981, \apj, 248, 321

\bibitem[{{Lin} \& {Pringle}(1987)}]{lin87}
{Lin}, D.~N.~C., \& {Pringle}, J.~E. 1987, \mnras, 225, 607

\bibitem[{{Lin}(2014)}]{lin14c}
{Lin}, M.-K. 2014, \apj, 790, 13

\bibitem[{{Lin} \& {Youdin}(2015)}]{lin15}
{Lin}, M.-K., \& {Youdin}, A.~N. 2015, \apj, 811, 17

\bibitem[{{Lodato} \& {Clarke}(2011)}]{lodato11}
{Lodato}, G., \& {Clarke}, C.~J. 2011, \mnras, 413, 2735

\bibitem[{{Lodato} \& {Rice}(2004)}]{lodato04}
{Lodato}, G., \& {Rice}, W.~K.~M. 2004, \mnras, 351, 630

\bibitem[{{Lodato} \& {Rice}(2005)}]{lodato05}
---. 2005, \mnras, 358, 1489

\bibitem[{{Lynden-Bell} \& {Pringle}(1974)}]{lynden-bell74}
{Lynden-Bell}, D., \& {Pringle}, J.~E. 1974, \mnras, 168, 603

\bibitem[{{Mamatsashvili} \& {Rice}(2010)}]{mamat10}
{Mamatsashvili}, G.~R., \& {Rice}, W.~K.~M. 2010, \mnras, 406, 2050

\bibitem[{{Martin} \& {Lubow}(2011)}]{martin11}
{Martin}, R.~G., \& {Lubow}, S.~H. 2011, \apjl, 740, L6

\bibitem[{{Meru} \& {Bate}(2011)}]{meru11}
{Meru}, F., \& {Bate}, M.~R. 2011, \mnras, 411, L1

\bibitem[{{Meru} \& {Bate}(2012)}]{meru12}
---. 2012, \mnras, 427, 2022

\bibitem[{{Mohandas} \& {Pessah}(2015)}]{mohandas15}
{Mohandas}, G., \& {Pessah}, M.~E. 2015, ArXiv e-prints, arXiv:1510.02729

\bibitem[{{Paardekooper}(2012)}]{paardekooper12}
{Paardekooper}, S.-J. 2012, \mnras, 421, 3286

\bibitem[{{Paardekooper} {et~al.}(2011){Paardekooper}, {Baruteau}, \&
  {Meru}}]{paardekooper11b}
{Paardekooper}, S.-J., {Baruteau}, C., \& {Meru}, F. 2011, \mnras, 416, L65

\bibitem[{{Papaloizou} \& {Savonije}(1991)}]{papaloizou91}
{Papaloizou}, J.~C., \& {Savonije}, G.~J. 1991, \mnras, 248, 353

\bibitem[{{Papaloizou} \& {Lin}(1989)}]{papaloizou89}
{Papaloizou}, J.~C.~B., \& {Lin}, D.~N.~C. 1989, \apj, 344, 645

\bibitem[{{Plume} {et~al.}(1997){Plume}, {Jaffe}, {Evans}, {Martin-Pintado}, \&
  {Gomez-Gonzalez}}]{plume97}
{Plume}, R., {Jaffe}, D.~T., {Evans}, N.~J., I., {Martin-Pintado}, J., \&
  {Gomez-Gonzalez}, J. 1997, \apj, 476, 730

\bibitem[{{Rafikov}(2015)}]{rafikov15}
{Rafikov}, R.~R. 2015, \apj, 804, 62

\bibitem[{{Rice} {et~al.}(2011){Rice}, {Armitage}, {Mamatsashvili}, {Lodato},
  \& {Clarke}}]{rice11}
{Rice}, W.~K.~M., {Armitage}, P.~J., {Mamatsashvili}, G.~R., {Lodato}, G., \&
  {Clarke}, C.~J. 2011, \mnras, 418, 1356

\bibitem[{{Rice} {et~al.}(2012){Rice}, {Forgan}, \& {Armitage}}]{rice12}
{Rice}, W.~K.~M., {Forgan}, D.~H., \& {Armitage}, P.~J. 2012, \mnras, 420, 1640

\bibitem[{{Rice} {et~al.}(2005){Rice}, {Lodato}, \& {Armitage}}]{rice05}
{Rice}, W.~K.~M., {Lodato}, G., \& {Armitage}, P.~J. 2005, \mnras, 364, L56

\bibitem[{{Schmit} \& {Tscharnuter}(1995)}]{schmit95}
{Schmit}, U., \& {Tscharnuter}, W.~M. 1995, Icarus, 115, 304

\bibitem[{{Shakura} \& {Sunyaev}(1973)}]{shakura73}
{Shakura}, N.~I., \& {Sunyaev}, R.~A. 1973, \aap, 24, 337

\bibitem[{{Shi} \& {Chiang}(2014)}]{shi14}
{Shi}, J.-M., \& {Chiang}, E. 2014, \apj, 789, 34

\bibitem[{{Shlosman} \& {Begelman}(1987)}]{shlosman87}
{Shlosman}, I., \& {Begelman}, M.~C. 1987, \nat, 329, 810

\bibitem[{{Shu}(1970)}]{shu70}
{Shu}, F.~H. 1970, \apj, 160, 99

\bibitem[{{Shu}(1977)}]{shu77}
---. 1977, \apj, 214, 488

\bibitem[{{Shu}(1984)}]{shu84}
{Shu}, F.~H. 1984, in IAU Colloq. 75: Planetary Rings, ed. R.~{Greenberg} \&
  A.~{Brahic}, 513--561

\bibitem[{{Stamatellos} \& {Whitworth}(2009)}]{stam09}
{Stamatellos}, D., \& {Whitworth}, A.~P. 2009, \mnras, 392, 413

\bibitem[{{Sterzik} {et~al.}(1995){Sterzik}, {Herold}, {Ruder}, \&
  {Willerding}}]{sterzik95}
{Sterzik}, M.~F., {Herold}, H., {Ruder}, H., \& {Willerding}, E. 1995, P\& SS,
  43, 259

\bibitem[{{Takahashi} \& {Inutsuka}(2014)}]{takahashi14}
{Takahashi}, S.~Z., \& {Inutsuka}, S.-i. 2014, \apj, 794, 55

\bibitem[{{Toomre}(1964)}]{toomre64}
{Toomre}, A. 1964, \apj, 139, 1217

\bibitem[{{Turner} {et~al.}(2014){Turner}, {Fromang}, {Gammie}, {Klahr},
  {Lesur}, {Wardle}, \& {Bai}}]{turner14}
{Turner}, N.~J., {Fromang}, S., {Gammie}, C., {et~al.} 2014, Protostars and
  Planets VI, 411

\bibitem[{{Ward}(2000)}]{ward00}
{Ward}, W.~R. 2000, {On Planetesimal Formation: The Role of Collective Particle
  Behavior} (University of Arizona Press), 75--84

\bibitem[{{Willerding}(1992)}]{willerding92}
{Willerding}, E. 1992, Earth Moon and Planets, 56, 173

\bibitem[{{Youdin}(2005)}]{youdin05}
{Youdin}, A.~N. 2005, ArXiv Astrophysics e-prints, astro-ph/0508659

\bibitem[{{Youdin}(2011)}]{youdin11}
---. 2011, \apj, 731, 99

\bibitem[{{Young} \& {Clarke}(2015)}]{young15}
{Young}, M.~D., \& {Clarke}, C.~J. 2015, \mnras, 451, 3987

\bibitem[{{Young} \& {Clarke}(2016)}]{young16}
---. 2016, \mnras, 455, 1438

\end{thebibliography}
